\documentclass[fleqn,usenatbib]{mnras}

\usepackage{newtxtext,newtxmath}
\usepackage[T1]{fontenc}

\DeclareRobustCommand{\VAN}[3]{#2}
\let\VANthebibliography\thebibliography
\def\thebibliography{\DeclareRobustCommand{\VAN}[3]{##3}\VANthebibliography}

\usepackage{graphicx}	% Including figure files
\usepackage{amsmath}	% Advanced maths commands
\usepackage{balance}
\usepackage{subcaption}
\usepackage{sidecap}
\usepackage{lipsum}
\usepackage{upgreek}
\usepackage{multirow}
\usepackage{xcolor}

\newcommand{\pcmm}{\,cm$^{-2}$}	% per cm-squared
\newcommand{\pcmmm}{\,cm$^{-3}$}	% per cm-cubed
\newcommand{\mic}{$\upmu$m}
\newcommand{\msun}{M$_\odot$}

\newcommand{\kms}{km$\,$s$^{-1}$}
\newcommand{\tast}{$T_A^*$}
\newcommand{\mtast}{T_A^*}
\newcommand{\tmb}{$T_\mathrm{mb}$}
\newcommand{\mtmb}{T_\mathrm{mb}}

\newcommand{\fw}{{\sc FellWalker}}
\newcommand{\fwhires}{\texttt{fwhires}}
\newcommand{\fwlores}{\texttt{fwlores}}
\newcommand{\fwchimps}{\texttt{fwchimps}}

\newcommand{\rgc}{R_\mathrm{GC}}
\newcommand{\rsig}{R_\sigma}
\newcommand{\req}{R_\mathrm{eq}}
\newcommand{\etarad}{\eta_R}
\newcommand{\vlsr}{$v_\mathrm{LSR}$}

\newcommand{\twco}{$^{12}$CO}
\newcommand{\tco}{$^{13}$CO}
\newcommand{\ceo}{C$^{18}$O}
\newcommand{\xco}{$X_\mathrm{CO}$}
\newcommand{\xtwco}{$X_\mathrm{^{12}CO\,(3-2)}$}
\newcommand{\xtco}{$X_\mathrm{^{13}CO\,(3-2)}$}
\newcommand{\xunits}{\pcmm\,(K\,\kms)$^{-1}$}

\newcommand{\xtwcounresolved}{$4.0\times10^{20}$\xunits}

\newcommand{\xtcoresolved}{$4.3\times10^{21}$\xunits}
\newcommand{\xtcounresolved}{$5.3\times10^{21}$\xunits}

\newcommand{\red}[1]{{\color{red}{}#1}} % Red text
\defcitealias{PlanckCollaboration+20}{Planck Collaboration 2020}

\title[PAMS: The Perseus Arm Molecular Survey]{PAMS: The Perseus Arm Molecular Survey -- I. Survey description and first results}

\author[A.~J.~Rigby et al.]{Andrew~J.~Rigby\thanks{E-mail:
a.j.rigby@leeds.ac.uk}$^{,1}$, 
Mark~A.~Thompson$^{1}$,
David~J.~Eden$^{2, 3}$,
Toby~J.~T.~Moore$^{4}$,
Mubela~Mutale$^{1}$,
\newauthor
Nicolas~Peretto$^{5}$,
Rene~Plume$^{6}$,
James~S.~Urquhart$^{7}$,
Gwenllian~M.~Williams$^{8}$,
and Malcolm~J.~Currie$^{9}$.\\
$^{1}$School of Physics and Astronomy, University of Leeds, Leeds LS2 9JT, UK\\
$^{2}$Physics Department, University of Bath, Claverton Down, Bath BA2 7AY, UK\\
$^{3}$Armagh Observatory and Planetarium, College Hill, Armagh BT61 9DB, UK\\
$^{4}$Astrophysics Research Institute, Liverpool John Moores University, Liverpool Science Park, 146 Brownlow Hill, Liverpool L3 5RF, UK\\
$^{5}$Cardiff Hub for Astrophysics Research \& Technology, School of Physics \& Astronomy, Cardiff University, Queens Buildings, Cardiff CF24 3AA, UK\\
$^{6}$Department of Physics and Astronomy, University of Calgary, 2500 University Drive NW, Calgary, Alberta T2N 1N4, Canada\\
$^{7}$Centre for Astrophysics and Planetary Science, University of Kent, Canterbury CT2 7NH, UK\\
$^{8}$Department of Physics, Aberystwyth University, Penglais, Aberystwyth, Ceredigion, SY23 3BZ, UK\\
$^{9}$RAL Space, STFC Rutherford Appleton Laboratory, Didcot OX11 0QX, UK.
}

\date{Accepted XXX. Received YYY; in original form ZZZ}

\pubyear{2024}

\begin{document}
% \tableofcontents
\label{firstpage}
\pagerange{\pageref{firstpage}--\pageref{lastpage}}
\maketitle

% Abstract of the paper: 250 words max. The abstract should briefly describe the aims, methods, and main results of the paper. It should be a single paragraph not more than 250 words (200 words for Letters). No references should appear in the abstract.
\begin{abstract}
The external environments surrounding molecular clouds vary widely across galaxies such as the Milky Way, and statistical samples of clouds are required to understand them. We present the Perseus Arm Molecular Survey (PAMS), a James Clerk Maxwell Telescope (JCMT) survey combining new and archival data of molecular-cloud complexes in the outer Perseus spiral arm in \twco, \tco, and \ceo\ ($J$=3--2). With a survey area of $\sim$8\,deg$^2$, PAMS covers well-known complexes such as W3, W5, and NGC\,7538 with two fields at $\ell \approx 110\degr$ and $\ell \approx 135\degr$. PAMS has an effective resolution of 17 arcsec, and rms sensitivity of $\mtmb = 0.7$--1.0 K in 0.3\,\kms\ channels. Here we present a first look at the data, and compare the PAMS regions in the Outer Galaxy with Inner Galaxy regions from the CO Heterodyne Inner Milky Way Plane Survey (CHIMPS). By comparing the various CO data with maps of H$_2$ column density from \emph{Herschel}, we calculate representative values for the CO-to-H$_2$ column-density $X$-factors, which are \xtwco$\,=4.0\times10^{20}$ and \xtco$\,=4.0\times10^{21}$\xunits\ with a factor of 1.5 uncertainty. We find that the emission profiles, size--linewidth, and mass--radius relationships of \tco-traced structures are similar between the Inner and Outer Galaxy. Although PAMS sources are slightly more massive than their Inner Galaxy counterparts for a given size scale, the discrepancy can be accounted for by the Galactic gradient in gas-to-dust mass ratio, uncertainties in the $X$-factors, and selection biases. We have made the PAMS data publicly available, complementing other CO surveys targeting different regions of the Galaxy in different isotopologues and transitions.
\end{abstract}
% 246 words

% Select between one and six entries from the list of approved keywords.
% Don't make up new ones.

%%%%%%%%%%%%%%%%%%%%%%%%%%%%%%%%%%%%%%%%%%%%%%%%%%

%%%%%%%%%%%%%%%%% BODY OF PAPER %%%%%%%%%%%%%%%%%%

\section{Introduction} \label{sec:introduction}

Star clusters form within regions of giant molecular clouds (GMCs) where self-gravity is able to overcome opposing physical processes such as thermal pressure, turbulence, and magnetic fields. There are many properties within the interstellar medium (ISM) of a galaxy that vary from one location to another, and these environmental differences could reasonably be expected to leave an imprint upon the process of star formation. For example, in the Milky Way, there are several key differences between the Inner and Outer Galaxy, which we define here as the regions inside and outside of the Sun's orbit with a Galactocentric radius of $\rgc = 8.15 \pm 0.15$\,kpc \citep{Reid+19}, respectively. The molecular-to-atomic gas ratio, drops from a value of $f_\mathrm{mol}\approx 1$ in the central molecular zone (CMZ, $\rgc < 0.5$\,kpc) at the centre of the Galaxy, to $f_\mathrm{mol} \approx 0.1$ at $\rgc = 10$\,kpc \citep{Sofue+Nakanishi16}. The strength of the interstellar radiation field \citep{Popescu+17}, dust temperature \citep{Marsh+17}, metallicity \citep{Luck+Lambert11}, and the ratio of solenoidal to compressive turbulence within molecular clouds \citep{Rani+22} all decrease with $\rgc$. 

The CMZ is a significantly different star-forming environment from the Galactic disc, with a star-formation rate (SFR) an order of magnitude lower than would be expected for the same surface density of molecular gas across a galactic disc \citep[e.g.][]{Longmore+13a, Barnes+17}. The low star-formation efficiency (SFE) in the CMZ is accompanied by very large line-of-sight velocity dispersions, and a size--linewidth relationship that is steeper than typically found in the Galactic disc \citep{Kauffmann+17a}. These results are generally interpreted as signatures of elevated levels of turbulence that suppress the SFR. \citet{Federrath+16} found that turbulent motions within the CMZ cloud G0.253$+$0.016 are consistent with being dominated by solenoidal modes, which are expected to inhibit star formation (and, conversely, turbulence that is dominated by compressive modes are expected to promote star formation). The authors speculate that shear is responsible for the strong solenoidal modes in the CMZ, caused by the differential rotation of gas that is stronger towards the centres of galaxies.  

It may then be expected that as shear decreases, moving from the interior to the exterior of the Galaxy, that solenoidal modes of turbulence become less important. Indeed, \citet{Rani+22} report a negative gradient in the ratio of power in solenoidal to compressive turbulence with Galactocentric radius ($\rgc$) for molecular clouds from CHIMPS \citep{Rigby+16}. They found that this is also accompanied by a rise in SFE, indicating that the mode of turbulence may play a role in determining SFE across the Galaxy. While the molecular clouds within the CHIMPS survey cover a relatively wide range of $\rgc$, from 4--12\,kpc, the number of sources beyond $\rgc > 8$\,kpc is relatively small, and the spatial resolution is rather limited due to the large heliocentric distances of $d > 12$\,kpc that result from the survey field covering a limited range of low Galactic longitudes (roughly 28\degr to 46\degr).

The Outer Galaxy presents a different star-forming environment from the Inner Galaxy and the CMZ. In addition to the various gradients mentioned above that produce different conditions in the Outer Galaxy, dynamical features also differ in important ways. The interval between the passage of spiral arms is relatively long, and the spiral arms themselves are wider \citep{Reid+19}. The spiral structure itself is less regular, with large deviations from logarithmic-spirals. The corotation radius is located at $\rgc \sim$7\,kpc, outside of which the spiral arm pattern speed exceeds the circular velocity of the gas and stars, and the outer Lindblad resonance is located at $\rgc \sim$11\,kpc \citep{Clarke+Gerhard22}. Although spiral arms do not appear to substantially alter star formation in the Inner Galaxy \citep[e.g.][]{Moore+12,Eden+13,Eden+15,Urquhart+21,Colombo+22,Querejeta+24}, these dynamical effects may still have an impact (perhaps indirectly) upon the star-formation process in the Outer Galaxy, where the surface density of molecular clouds and their complexes is much lower.

Large-scale and unbiased surveys of dust continuum -- such as The Apex Telescope Large Area Survey of the Galaxy, \citep[ATLASGAL;][]{Schuller+09} and the \emph{Herschel} infrared Galactic Plane Survey \citep[Hi-GAL;][]{Molinari+16} -- and CO -- such as the \tco\ (1--0) Galactic Ring Survey \citep[GRS;][]{Jackson+06}, CHIMPS \citep{Rigby+16} in \tco\ (3--2), the CO High-Resolution Survey \citep[COHRS;][]{Dempsey+13, Park+23} in \twco\ (3--2), and the Structure, Excitation, and Dynamics of the Inner Galactic Interstellar Medium \citep[SEDIGISM;][]{Schuller+17} survey in \tco\ (2--1) -- have advanced our understating of many of the Galaxy-scale phenomena listed above in the Inner Galaxy. 

However, the relative sparsity of sources in the Outer Galaxy (i.e. the number of molecular clouds per unit angular area) mean that it is more difficult to justify the expenditure of time on unbiased (i.e. blind) mapping in molecules rarer than \twco\ at moderate resolution. Survey data do exist in the Outer Galaxy: the FCRAO and Exeter-FCRAO CO Galactic Plane Surveys cover $55 \ge \ell \leq 195$ in \twco\ and \tco\ (1--0) with varying coverage in Galactic latitude at 46-arcsecond resolution \citep[][Brunt et al. in prep.]{Heyer+98,Wienen+22}; the Forgotten Quadrant survey covers longitudes of $220 < \ell < 240\degr$ with $2.5 < b < 0.0\degr$ \citep{Benedettini+20}; and the Milky Way Imaging Scroll Painting survey \citep[MWISP;][]{Su+19} covers this region within its staggering  $-10\degr \leq \ell \leq 250\degr$ and $|b| < 5\fdg2$ footprint in \twco, \tco, and \ceo\ (1--0) at 50-arcsecond resolution. The latest generation of surveys at $\lesssim 30$-arcsecond resolution is only now catching up with their Inner Galaxy counterparts: the FOREST unbiased Galactic Plane imaging survey with the Nobeyama 45-m telescope \citep[FUGIN;][]{Umemoto+17}, covers \twco, \tco, and \ceo\ (1--0) at $\sim$20-arcsecond resolution covers $198\degr \ge \ell \leq 236\degr$ in the third quadrant, which is now also partially ($215\degr \ge \ell \leq 225\degr$) covered by the CHIMPS2 survey (\citealt{Eden+20}). The CO Large Outer-Galaxy Survey (CLOGS; Eden et al. in preparation) is now extending an area adjoining the CHIMPS2 Outer Galaxy survey, primarily in \twco, but with follow-ups of bright sources in \tco\ and \ceo (3--2); and the Outer Galaxy High-Resolution Survey \citep[OGHReS;][]{Urquhart+24} is also mapping a large area of the third quadrant in the \twco\ and \tco\ (2--1) with a similar specification as SEDIGISM. In dust continuum, the SCUBA-2 Ambitious Sky Survey (SASSy; \citealt{Nettke+17}, Thompson et al. in preparation) and SASSy-Perseus (Thompson et al. in preparation) surveys cover $120\degr < \ell < 250\degr$ and $60\degr < \ell < 120\degr$, respectively, and have been able to map a very large area in 850\,\mic\ with a sensitivity comparable to that of ATLASGAL. 
% AJR: This section would benefit from a table.

The second Galactic quadrant ($90\degr \leq \ell \leq 180\degr$) is relatively unexplored at high angular resolution. In this paper we present the 17-arcsecond-resolution Perseus Arm Molecular Survey (PAMS), a survey of several Outer Galaxy star-forming regions in the second quadrant in the 3--2 rotational transition of \tco\ and \ceo, and incorporating new and archival \twco\ data that cover most of the PAMS regions. The observations are highly complementary to other CO surveys such as CHIMPS/2, COHRS, and CLOGS, and greatly bolster the available statistics of Outer Galaxy star-forming regions. The surveyed regions include some of the most famous star-forming regions within the Perseus arm in the Outer Galaxy, such as W3, W5, and NGC\,7538, which lie at $\rgc \approx 9.5$\,kpc. 

In Section~\ref{sec:observations} we describe the observations, and we present the data in Section~\ref{sec:data}. In Section~\ref{sec:results} we compare analyses of the CO-to-molecular-hydrogen column-density conversion factor, and basic molecular-cloud scaling relationships between the Outer Galaxy PAMS data and representative Inner Galaxy data from the CHIMPS survey. We discuss our findings and conclusions in Section~\ref{sec:conclusions}.

% =================================================================
% =================================================================
\section{Observations \& Data Reduction} \label{sec:observations}
% =================================================================
% =================================================================

% =========== Ideally group this figure and table together ========
\begin{figure*}
    % RMS maps generated from Make_RMS_maps.sh
    % Generated by Figure_Targets.py
    \centering
    \includegraphics[width=0.9\textwidth]{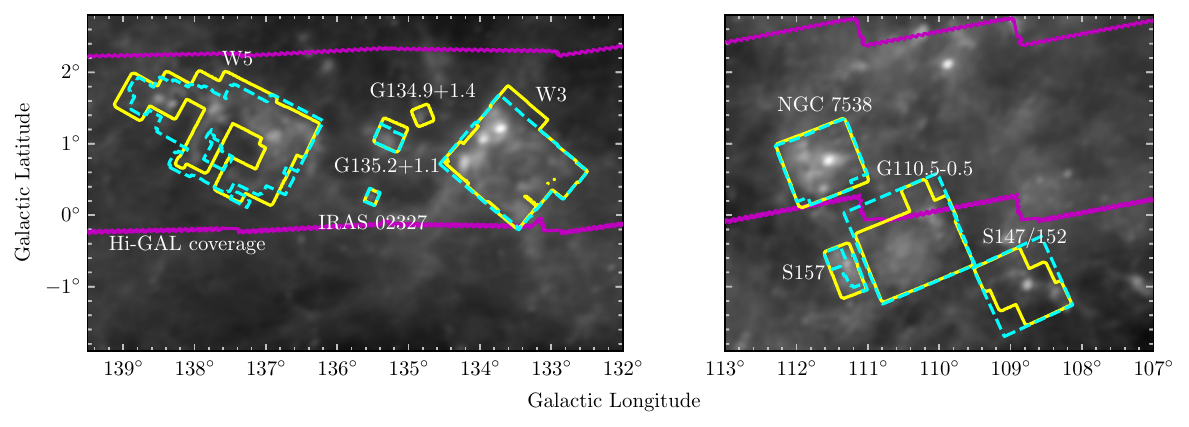}
    \caption{Illustration of the extent of each PAMS region overlaid on \emph{Planck} 857 GHz continuum imaging \citep{PlanckCollaboration+20}, which is displayed on a logarithmic intensity scale. \emph{Left panel}: the $\ell=137\degr$ region. \emph{Right panel}: the $\ell=111\degr$ region. The solid yellow and dashed cyan lines show the extent of the PAMS \tco\ and \ceo (3--2) data, and \twco\ (3--2) data, respectively. The extent of Hi-GAL 160 and 250\,\mic\ coverage is outlined in magenta.}
    \label{fig:targets}

\end{figure*}

\begin{table*}
% Coords generated by Get_coords.py
% Velocities from Get_veloities.py
% Table generated by Characterise_data.py
\centering
\caption{Details of the target regions, in reverse order of Galactic longitude: approximate field centres in Galactic coordinates, representative systemic velocity determined from a Gaussian fit to the mean \tco\ (3--2) spectrum, heliocentric distance and uncertainty, Galactocentric distance, total area of the PAMS observations, and the rms values of the various observations on the \tmb\ scale at the two different velocity resolutions.}
\begin{tabular}{@{\extracolsep{2pt}}cccccccccccccc@{}}
\hline
Region & $\ell$ & $b$ & $v_\mathrm{LSR}$ & $d$ & $\Delta d$ & $R_\mathrm{GC}$ & Area & \multicolumn{6}{c}{rms(\tmb)} \\
\cline{9-14}
\rule{0pt}{3ex}  
& & & & & & & & \multicolumn{2}{c}{$^{12}$CO} & \multicolumn{2}{c}{$^{13}$CO} & \multicolumn{2}{c}{C$^{18}$O} \\
\cline{9-10} \cline{11-12} \cline{13-14}
\rule{0pt}{3ex}  
 &  &  &  & &  & & & 0.3 & 0.5 & 0.3 & 0.5 & 0.3 & 0.5 \\
  &  &  &  & &  & & & \kms & \kms & \kms & \kms & \kms & \kms \\
 & ($\mathrm{{}^{\circ}}$) & ($\mathrm{{}^{\circ}}$) & (\kms) & (kpc) & (kpc) & (kpc) & (deg$^2$) &($\mathrm{K}$) &($\mathrm{K}$) & ($\mathrm{K}$) & ($\mathrm{K}$) & ($\mathrm{K}$) & ($\mathrm{K}$) \\ \hline
W5 & 137.8 & 1.3 & -38.6 & 1.96 & 0.04 & 9.69 & 2.42 & 1.01 & 0.75 & 0.66 & 0.51 & 0.83 & 0.69 \\
IRAS\,02327 & 135.5 & 0.3 & -43.1 & 1.96 & 0.04 & 9.65 & 0.03 & -- & 0.91 & 0.54 & 0.42 & 0.72 & 0.56 \\
G135.2+1.1 & 135.2 & 1.1 & -44.8 & 1.96 & 0.04 & 9.64 & 0.14 & -- & 0.60 & 0.81 & 0.63 & 1.05 & 0.79 \\
G134.9+1.4 & 134.8 & 1.4 & -40.6 & 1.96 & 0.04 & 9.63 & 0.06 & -- & -- & 0.90 & 0.70 & 1.15 & 0.97 \\
W3 & 133.5 & 0.7 & -45.4 & 1.96 & 0.04 & 9.60 & 1.97 & 0.76 & 0.59 & 0.61 & 0.47 & 0.84 & 0.65 \\
NGC\,7538 & 111.6 & 0.7 & -53.1 & 2.69 & 0.13 & 9.48 & 0.99 & -- & 0.35 & 0.68 & 0.53 & 0.89 & 0.73 \\
S157 & 111.3 & -0.8 & -43.4 & 3.38 & 0.15 & 9.89 & 0.26 & -- & 0.85 & 0.97 & 0.75 & 1.3 & 1.0 \\
G110.5-0.5 & 110.4 & -0.3 & -51.3 & 2.72 & 0.19 & 9.45 & 1.57 & 1.39 & 1.16 & 0.67 & 0.52 & 0.86 & 0.7 \\
S147/152 & 108.8 & -1.0 & -50.8 & 2.81 & 0.22 & 9.44 & 0.80 & 1.00 & 0.84 & 0.81 & 0.63 & 1.03 & 0.83 \\
\hline
Total &  &  &  &  &  &  & 8.24 & 1.00 & 0.72 & 0.68 & 0.53 & 0.88 & 0.71 \\
\hline
\end{tabular}
\label{tab:observations}
\end{table*}
% =======================================================================

The data presented here as PAMS are comprised of data cubes of the (3--2) transition of \twco, \tco, and \ceo\ for nine regions in the Outer Perseus spiral arm. Eight of those regions were originally targeted for PAMS in \tco\ and \ceo\ with observing campaigns in 2009--10, and we have supplemented these with archival \twco\ data, where available, and obtained new observations of G110 and S152. By incorporating further archival observations of W3, PAMS covers a total of nine regions, with almost complete coverage in the three isotopologues. The observations and data reduction are detailed below.

\subsection{\tco\ and \ceo\ (3--2) observations}
We conducted simultaneous basket-woven raster mapping observations of \tco\ and \ceo\ ($J$=3--2) at 330.588~GHz and 329.331~GHz using the Heterodyne Array Receiver Program and Auto-Correlation Spectral Imaging System \citep[HARP/ACSIS;][]{Buckle+09} at the 15-m James Clerk Maxwell Telescope (JCMT) on Mauna Kea, Hawaii. The observations, taken in 2009--10, consist of a series of tiles of up to 1020$\times$1020 arcsec in size as part of projects M09BU04 and M10BU08. The targets are listed in Table~\ref{tab:observations}, and the field extents are illustrated upon \emph{Planck} 857 GHz continuum maps \citepalias{PlanckCollaboration+20} in Fig.~\ref{fig:targets}. For each tile, these observations took the form of two sets of position-switched scans at right angles to each other with a quarter array (29.1 arcsec) shift between each scan in a given direction. The same reference positions were used for each tile in a given region, which were checked to ensure they were free of contamination. The 250-MHz bandwidth correlator setting was used with 4096 channels, resulting in a native spectral resolution of $\sim$0.06~\kms. The native angular resolution of the JCMT at 330 GHz is 15 arcsec.

Following standard practice at JCMT, pointing was checked between observations, for which the uncertainty is estimated to be 2 arcsec in both azimuth and elevation, resulting in a 3 arcsec radial uncertainty. Calibration was performed using the three-load chopper-wheel method \citep{Kutner+Ulich81} during the observations, with which the intensity of the spectra are placed on the corrected antenna temperature (\tast) scale. Spectral standards are also monitored throughout observations, and peak and integrated flux densities are generally found to be accurate to within 10 per cent.
 
\subsection{\twco\ (3--2) observations} \label{sec:newobs}
New observations of \twco\ (3--2) at 345.796~GHz for the G110 and S152 regions were obtained in the summer of 2024 through Programs M23BN003 (PI: Parsons) and S24BP004 (PI: Eden). The setup was similar to the \tco\ and \ceo\ observations except for the following differences: the backend (ACSIS) was configured to use 8192 250-MHz-wide channels, all individual tiles were 1320$\times$1320 arcsec in size, and half-array (58.2 arcsec) spacing was used between rows. The native angular resolution of JCMT at 345 GHz is 14 arcsec.

\begin{figure*}
    \centering
    % Figure_NGC7538_mcf.py
    \includegraphics[width=0.8\textwidth]{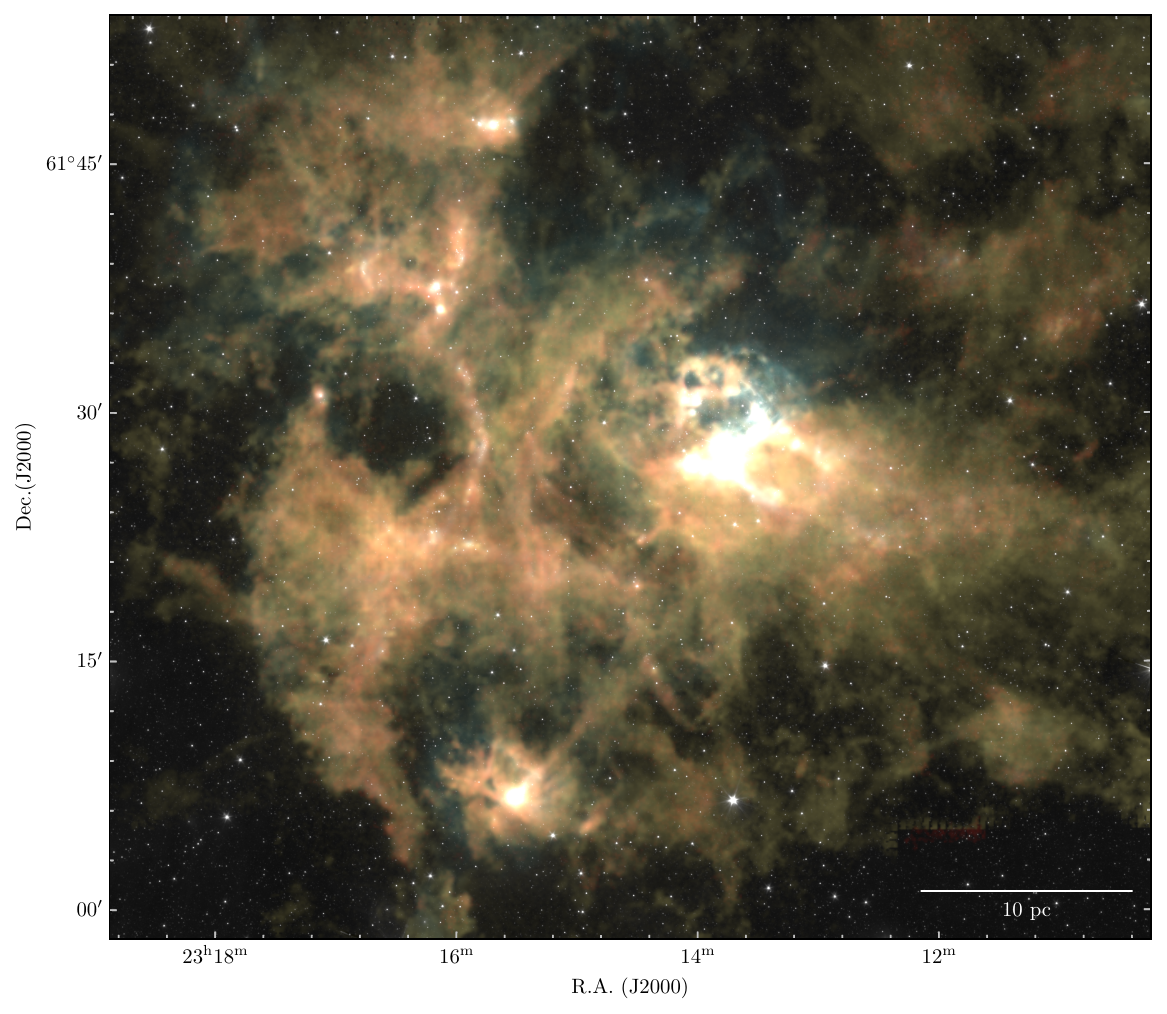}
    \caption{Four colour composite image of NGC\,7538. \emph{Spitzer}/GLIMPSE 4.5\,\mic\ \citep{Benjamin+03}, \emph{Herschel}/HOBYS 70\,\mic\ \citep{Motte+10}, archival JCMT \twco\ (3--2) integrated intensity, and JCMT/PAMS \tco\ (3--2) integrated intensity images are shown in white, cyan, yellow, and red channels, respectively. The integrated intensity CO and \tco\ (3--2) images have been masked as described in Section~\ref{sec:sourceextraction}, and are individually displayed in Fig.~\ref{fig:NGC7538}.}
    \label{fig:mcf}
\end{figure*}

\subsection{Archival observations}
We have incorporated further $^{12}$CO ($J$=3--2) data from the JCMT archives covering five of the regions: G135.2+1.1, IRAS\,02327, S157, NGC\,7538, and W5 have almost complete coverage and, while G110 and S152 had partial coverage in the archive, those data were superseded by our new observations. G134.9+1.4 is the only region that has no \twco\ (3--2) coverage. The archival data used were M07AU08 (PI: S. Lumsden), M07BH45B, M08BH15, and M09BC12 (PI: J. Williams), M09BN07 (PI: M. Hogerheijde), M09BH09C (PI: J. Di Francesco) and M10BC04 (PI: M. Reid). The NGC\,7538 and W5 data were previously published in \citet{Fallscheer+13} and \citet{Ginsburg+11}, respectively, while outflows within G110, S152, and S157 formed part of the sample of \citet{Maud+15}.

Additional \twco, \tco, and \ceo\ (3--2) observations of the W3 complex were incorporated from observing programs M06BU21, M07BH17B, M08BU24 (PI: Moore). The W3 data were originally presented in \citep{Polychroni+12}. Some of these observations date from earlier in the lifetime of the then recently commissioned HARP instrument and have a slightly different observing format, most notably that the scans are not all basket-woven.

\subsection{Data reduction}
Data reduction was performed using \textsc{orac-dr} \citep{Jenness+15}, which is built on the Starlink \citep{Currie+14} packages \textsc{kappa} \citep{Currie+Berry13}, \textsc{cupid} \citep{Berry+07}, and \textsc{smurf} \citep{Chapin+13}, with which we used the \texttt{REDUCE\_SCIENCE\_NARROWLINE} recipe. We give the template reduction parameters in Appendix~\ref{app:recpars}. The \tco\ and \ceo\ data were regridded onto the 6-arcsec pixels using an 8-arcsec FWHM Gaussian smooth, meaning that the reduced data cubes have an effective angular resolution of 17.2 arcsec. In the vast majority of cases, first-order polynomials were used for baseline subtraction, although fourth-order fits were adopted in some cases. The final cubes were regridded onto 0.3\,\kms-wide velocity channels as our primary data products, but a second version was also produced with 0.5\,\kms\ channels to increase compatibility with CHIMPS \citep{Rigby+16} and CHIMPS2 \citep{Eden+20}. 

The \twco\ (3--2) data were reduced in an almost identical way using {\sc orac-dr}, differing only in the use of a 9-arcsec FWHM Gaussian smooth during the regridding, which results in data cubes whose effective angular resolution matches the 17.2-arcsecond resolution of the \tco\ and \ceo\ data cubes. Some of the CO data were observed with ACSIS configured with a 1000 MHz bandwidth rather than 250 MHz, which provides a native resolution of 0.42\,\kms, and so the CO data were rebinned to 0.5\,\kms\ velocity channels throughout for consistency, with the exception of G110, S152, and part of W3, for which the 0.3\,\kms\ cubes are also available. 

Mosaics of each of the regions were produced using {\sc kappa:wcsmosaic} with inverse-variance weighting and the \texttt{sincsinc} interpolation kernel. The cubes were astrometrically matched such that the \ceo\ and \twco\ mosaics share the same pixel grid and size as their \tco\ counterparts. The individual reduced data cubes are on the corrected antenna temperature (\tast) scale, and we converted the larger mosaics to main beam brightness temperature scale by dividing by the main beam efficiency, $\mtmb = \mtast / \eta_\mathrm{mb}$, where $ \eta_\mathrm{mb}$=0.72 at 330 GHz, or $\eta_\mathrm{mb}=0.61$ at 345 GHz \citep{Buckle+09}. In Fig.~\ref{fig:mcf} we display a combined view\footnote{Colour image created using the \href{https://github.com/pjcigan/multicolorfits}{\texttt{multicolorfits}} Python package.} of the \twco\ (3--2) emission alongside the PAMS \tco\ (3--2) emission for NGC\,7538, highlighting the utility of the combined data sets. The regions of orange emission trace the highest column densities of CO, while those of yellow emission trace the diffuse envelope of the region.

\subsection{Ancillary data}

We make use of the data from the CHIMPS survey \citep{Rigby+16} directly in Section~\ref{sec:XCO}, and use the clump catalogue from \citet{Rigby+19} in Section~\ref{sec:properties}.

We also use 160 and 250\,\mic\ data from Hi-GAL \citep{Molinari+16} to construct maps of H$_2$ column density and dust temperature in Section~\ref{sec:XCO}. These data were reduced using {\sc UniMap} -- the University of Rome and IAPS Map Maker -- version 5.1.0 (Molinari et al. in prep), which include de-striping, and the application of photometric offsets determined by comparison to \emph{Planck} data.
 
%%%%%%%%%%%%%%%%%%%%%%%%%%%%%%%%%%%%%%%%%%%%%%%%%%%%%%%%%%%%%%%%%%%%%%%%%%%%%%

\section{The Data} \label{sec:data}

\subsection{Data quality} \label{sec:dataquality}

\begin{figure*}
  \centering
    \centering
    % Characterise_Data.py
    \includegraphics[width=0.9\linewidth]{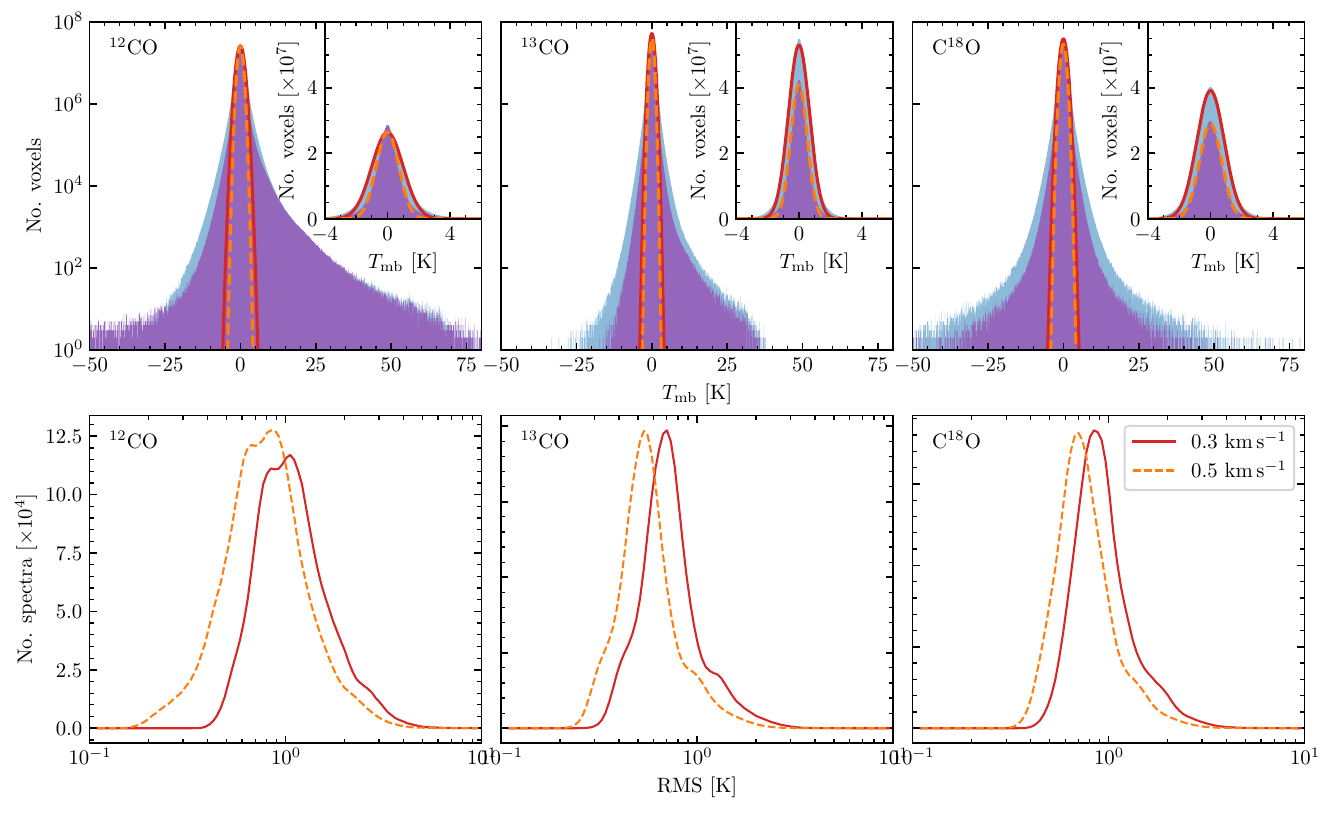}
    \caption{Data quality for \twco, \tco, and \ceo\ (3--2) emission in PAMS. \emph{Top row}: Histograms of voxel values in the three transitions with 0.05\,K-wide bins. The light and dark-shaded distributions correspond to the 0.3 and 0.5\,\kms\ channel-width data, respectively. The solid and dashed curves give the best Gaussian fits to the 0.3 and 0.5\,\kms\ channel-width data, respectively. The inset axes show the same distributions with a linear y-axis. \emph{Bottom row}: Histograms of rms values for the spectra in the \twco, \tco, and \ceo\ $J$=3--2 PAMS data using bins of width 0.02 dex. The 0.3\,\kms-binned and 0.5\,\kms-binned data are shown in solid and dashed lines, respectively.}
    \label{fig:voxeldist}    
    \label{fig:RMSdist}
\end{figure*}

\begin{figure*}
    \centering
    % Characterise_Data.py
    \includegraphics[width=\textwidth]{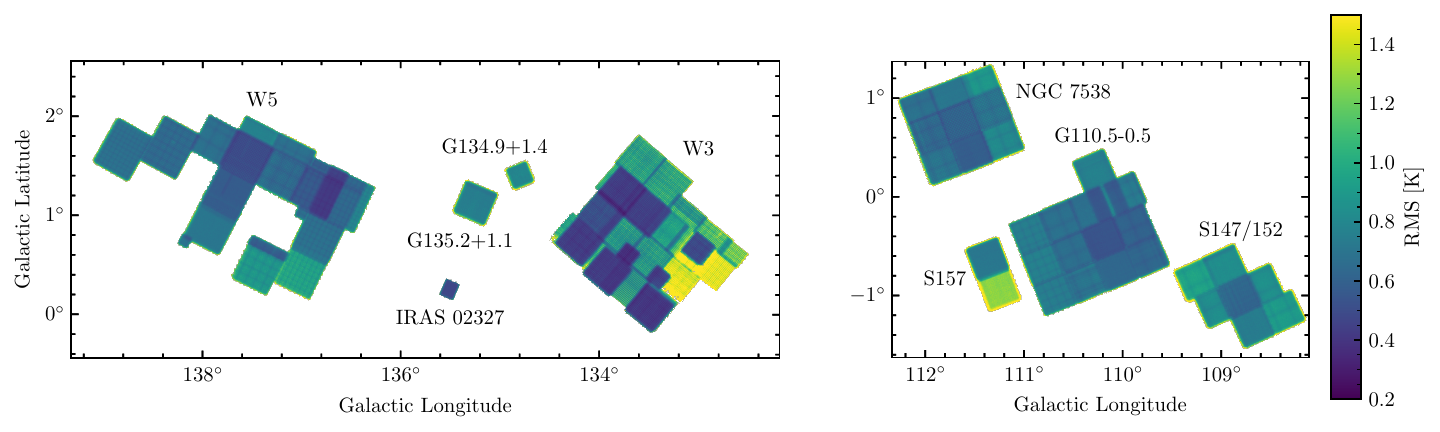}
    \caption{RMS maps for \tco\ (3--2) with 0.3\,\kms-wide channels for the $\ell=137\degr$ (\emph{left panel}) and $\ell=110\degr$ (\emph{right panel}) regions.}
    \label{fig:RMSmaps}
\end{figure*}

% ================================================================

Fig.~\ref{fig:voxeldist} shows the distributions of voxels (i.e. three-dimensional pixel) values for the entirety of the PAMS \twco, \tco, and \ceo\ (3--2) data, and in both the 0.3- and 0.5-\kms\ velocity channel-width variants. To determine a global rms value for each isotopologue, we fitted a normal distribution centred on a value of zero, and recovered rms values of 1.00 (0.72), 0.68 (0.53), and 0.88 (0.71)\,K in \tmb\ for \twco, \tco, and \ceo\ at 0.3 (0.5) \kms-resolution, respectively. The RMS-equivalent H$_2$ column densities are $\sim6\times10^{18}$, $3\times10^{20}$, and $3\times10^{21}$ \pcmm\ for \twco, \tco, and \ceo\ (3--2), respectively, assuming optically thin emission with an excitation temperature of 10\,K. For these figures, we have assumed abundance ratios of $8.5 \times 10^{-5}$ for \twco/H$_2$ \citep{Frerking+82}, and 79 and 596 for \twco/\tco\ and \twco/\ceo, respectively, from \citet{Wilson+Rood94} evaluated at $R_\mathrm{GC}=9.5$\,kpc. In all cases, the data are not perfectly normally distributed; the global rms values arise from the central limit theorem when combining a different distribution for each tile within each region. We display the values for each individual region in Table~\ref{tab:observations}. In Fig.~\ref{fig:voxeldist}, the logarithmic $y$-axes allow the non-Gaussian wings of the distributions to be seen most clearly, while the linear $y$-axes in the inset figures demonstrate the overall normal distributions. In \twco\ and \tco, the distributions show significant positive wings associated with the emission. This is less obvious in \ceo\ due to the lower relative abundance of the isotopologue, resulting in a much lower detection rate. The noise levels vary from tile to tile as a result of different observing conditions: zenith opacity (i.e. precipitable water vapour), target elevation, and the number of functioning receptors on HARP. The tile-to-tile noise variations can be seen in Fig.~\ref{fig:RMSmaps}, and we display the distributions of noise values for the various PAMS data sets in Fig.~\ref{fig:RMSdist}.

\subsection{Emission maps}

\begin{figure*}
    % Figure_NGC7538.py
    \centering
    \includegraphics[width=\textwidth]{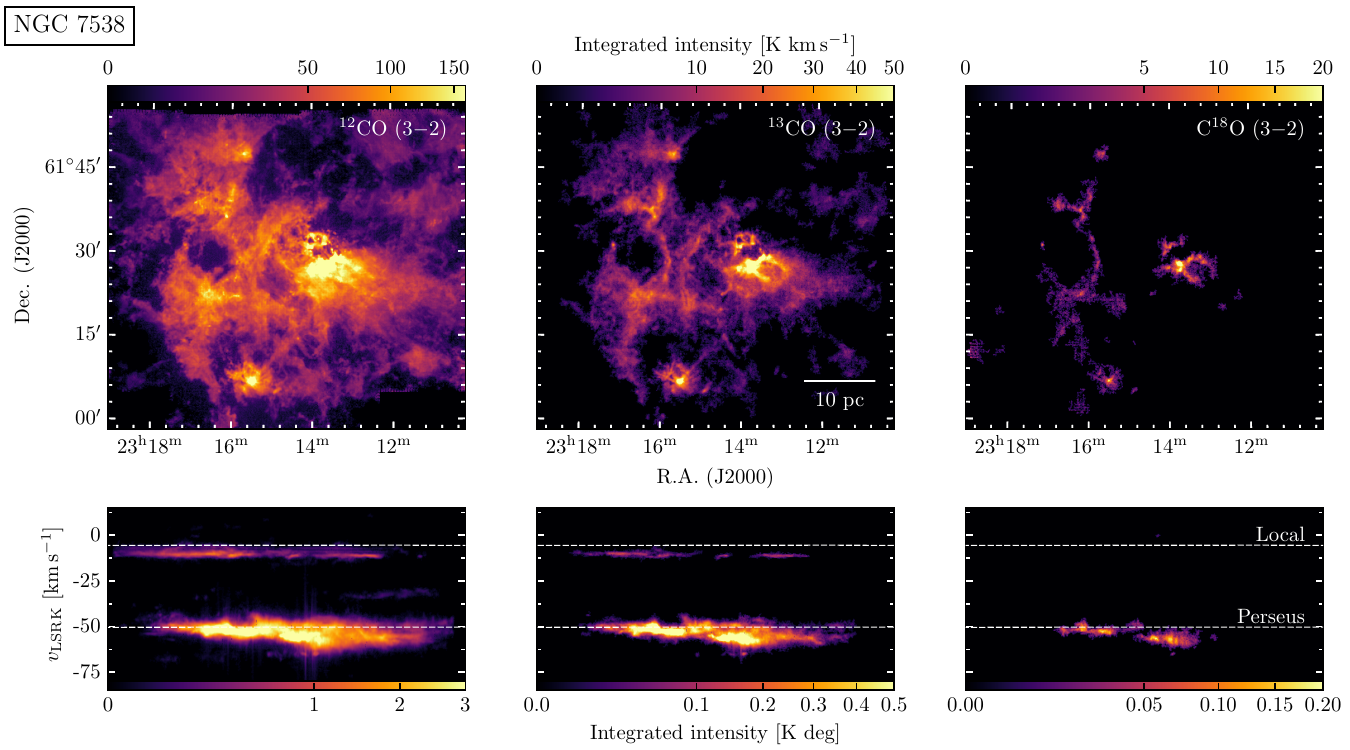}
    \caption{Integrated intensity images of \twco, \tco, and \ceo\ (3--2) for NGC\,7538, masked using \fw. \emph{Top row}: Images integrated over the velocity axis. \emph{Bottom row}: Images integrated over the declination axis overlaid with the loci of the spiral arms models of  \citet{Reid+19} that are present in this quadrant of the Galaxy and within the PAMS latitude range: the Local Arm, and the Perseus Arm.}
    \label{fig:NGC7538}
\end{figure*}

In Fig.~\ref{fig:NGC7538}, we present the observations of NGC\,7538, and equivalent figures for the other regions are shown in Appendix~\ref{app:regions}. We show both the moment 0 (velocity-integrated intensity) maps, and position-velocity maps (declination-integrated intensity) generation for each isotopologue of \twco. The images were first masked using the {\sc FellWalker} source-extraction software (discussed in Section~\ref{sec:sourceextraction}). Although the bulk of the emission from NGC\,7538 is contained within $-$70 to $-$40\,\kms, the region contains a number of outflows that extend beyond this range. The outflows are clearly visible in the position-velocity maps of \twco\ and, to a lesser extent, \tco. The field also contains a minor secondary emission component between $-$17 and $-$3\,\kms, which contains just 8 and 1 per cent of the total emission in \twco\ and \tco\ (3--2), respectively.

With the lowest \emph{effective} critical density, the \twco\ (3--2) emission traces the most diffuse component of the molecular cloud, which fills much of the field of view, and it is especially powerful for tracing outflows. By contrast, the \tco\ and \ceo\ (3--2) emission traces higher-density components; \twco\ is $\approx$ 80 and 600 times more abundant than \tco\ and \ceo, respectively, at a Galactocentric distance of 9.5\,kpc \citep{Wilson+Rood94}, and so the rarer isotopologues have substantially lower optical depths. \citet{Rigby+19} found that \tco\ (3--2) emission within molecular clouds in the inner Milky Way is predominantly $\tau < 1$, and only becomes optically thick towards the densest $\sim$1\,pc-scale clumps; only 3 per cent of clumps in the CHIMPS sample have mean optical depths greater than 1. NGC\,7538 and the other PAMS regions also reside substantially further out in the Galaxy than the clumps typically targeted by CHIMPS, and so optical depths are likely to be even lower for PAMS. \tco\ (3--2) emission is, therefore, expected to be a reasonably good tracer of H$_2$ column density where it is detected in these regions.

In Fig.~\ref{fig:NGC7538} we also overlay the loci of the models of the two spiral arms of \citet{Reid+19} that are present in this quadrant of the Galaxy, which fall within the PAMS latitude range. We clearly detect emission associated with the Local Arm in addition to the expected Perseus Arm -- with which the main PAMS regions are associated. The \citet{Reid+19} models are based upon trigonometric parallaxes of $\sim$200 high-mass star-forming regions, obtained by the Bar and Spiral Structure Legacy (BeSSeL) Survey\footnote{\url{http://bessel.vlbi-astrometry.org/}} and the Japanese VLBI Exploration of Radio Astrometry (VERA) project\footnote{\url{https://www.miz.nao.ac.jp/veraserver/}} which are known to a high level of accuracy. NGC\,7538 clearly resides within the Perseus arm, and the secondary emission component is consistent with a position in the Local Arm.

\subsection{Source extraction} \label{sec:sourceextraction}

A series of source extractions was performed upon the \tco\ (3--2) PAMS data using the \fw\ \citep{Berry15} algorithm, which is part of {\sc cupid} \citep{Berry+07}. \fw\ is a watershed clump-finding algorithm which segments our arrays of voxels into discrete clumps of emission, each of which contains a significant emission peak. {\sc FellWalker} assigns all voxels brighter than a threshold level, determined by the noise, to a single clump in this way. In addition to producing a catalogue, it also generates a mask which has the same dimensions as the input array, but in which each voxel value corresponds to the ID of a catalogued clump. We refer to these masks as `assignment masks' hereafter, and these may also be used to effectively remove the background from the data cubes, amongst other useful applications.

Source extraction was performed upon the signal-to-noise ratio (SNR) mosaics of each region. Initial tests showed that \fw\ was much more effective at locating the emission present within the data when running over the SNR cubes as opposed to the \tmb\ cubes, resulting in a smaller amount of emission remaining in the residual cubes. After an initial source search on the SNR cubes using {\sc cupid:findclumps}, the {\sc cupid:extractclumps} algorithm then uses the \fw-defined mask to extract information from the \tmb\ cubes, recalculating properties such as the intensity-weighted centroid and peak coordinate of each source, which is different in \tmb-space compared with SNR-space. A further advantage is that an extraction of sources on the SNR cubes reduces the instances of false-positive detections which can arise as a consequence of noisy spectra at the image edges or at the seams of the mosaics.

A total of three source extractions were run, with the following purposes:

\begin{enumerate}
    % NB: See Sensitivity_vs_resolution.pdf.
    \item \fwhires: Our primary source extraction, which was configured to locate objects on the `clump' size-scale, allowing localised levels of fragmentation.

    \item \fwlores: This setup essentially identifies the same pixels of emission as \fwhires, but was configured to retain the largest structures within each field.

    \item \fwchimps: A reference extraction was run with a set of parameters optimised to recover, as closely as possible, a catalogue that is consistent with the CHIMPS survey whose extraction was described in \citet{Rigby+16} and analysed in \citet{Rigby+19}.
\end{enumerate}

For both \fwhires\ and \fwlores, source extraction was performed on the 0.3-\kms-resolution mosaics, essentially giving the `best' case for our PAMS data, while the \fwchimps\ extraction was conducted on the 0.5\,\kms-resolution mosaics to facilitate a more direct comparison with CHIMPS. These source extractions were also performed after smoothing the data to a resolution of 22.0 arcsec, which was found to be the best compromise between retaining relatively high resolution, while improving the noise statistics. For these extractions, the minimum height of a peak to be included was set to SNR = 3, and the noise level was set to SNR = 1, meaning that all contiguous pixels down to the rms value are considered to be part of each source. Although the noise level (i.e. minimum voxel value to be included in a clump) is rather low, we found that higher levels for this parameter led to considerably more emission in the residuals. As always, flux boosting will be present in the low-SNR sources, and we stress that this extraction has been tuned to maximise the recovery of the emission, as opposed to catalogue robustness (i.e. minimizing false positives). The difference in the \fwhires\ and \fwlores\ was achieved by setting different values of \texttt{MinDip} which was set to an SNR value of 5 in the \fwhires\ case, and 1000 in the \fwlores\ case. By contrast, the \fwchimps\ extraction was configured to use the parameters described in \citep{Rigby+16}, with the exception that the \texttt{RMS} parameter was set to 1.7, reflecting the lower sensitivity of the CHIMPS data after smoothing to the same resolution of 27.4 arcsec that was originally used. One further difference in the process is that this extraction identifies sources at a resolution of 27.4 arcsec, but then extracts the parameters at the native resolution of 17.2 arcsec (15.2 arcsec in the original CHIMPS extraction). Both PAMS extractions identify and extract source parameters at the same resolution of 22.0 arcsec. We list the full set of \fw\ parameters used for each setup in Appendix~\ref{app:fwconfig}.

While \fw\ does not nominally contain any information about the hierarchical structure of the emission, our twin \fwhires\ and \fwlores\ allow some aspects of this to be recovered. Both extractions identify the same pixels of emission, and differ only in the assignment to catalogued structures. Because of this, we are able to assign every clump within the \fwhires\ extraction to a larger structure from the \fwlores\ extraction, and thus restore some information about the hierarchy. By contrast, the {\sc scimes} \citep{Colombo+15} algorithm is based upon the {\sc astrodendro} implementation of dendrograms \citep{Rosolowsky+08}, which identify substructures that are significant in terms of brightness and area based upon contour levels. While dendrograms also keep track of compact sources (identified as \emph{leaves} by analogy) inside larger substructures (\emph{branches}) of molecular clouds (\emph{trunks}), their individual values may differ compared with the equivalent {\sc FellWalker}-defined clumps, but statistically the two approaches return compatible results \citep{Rani+23}. In this paper, we use {\sc FellWalker} to enable a direct comparison with the properties of the clump population identified within the CHIMPS survey of \citet{Rigby+19} in the Inner Galaxy.

\begin{table*}
\centering
\caption{Information about sources extracted from the \fwhires\ source extraction. The columns give the IAU-compliant designation, the PAMS region, source ID in assignment cube, centroid longitude, centroid latitude, centroid velocity, velocity dispersion, equivalent radius, sum of \tco\ pixel values, peak \tco\ pixel value, peak signal-to-noise ratio, and ID of parent source in corresponding \fwlores\ extraction. The first five rows, and selected columns only are included here for illustrative purposes. The full catalogue, along with the full \fwlores\ and \fwchimps\ catalogues, are available in machine-readable format, as detailed in the Data Availability section.}
\begin{tabular}{cccccccccccc}
\hline
Designation & Region & ID & $\ell$ & $b$ & $v_\mathrm{lsr}$ & $\sigma(v_\mathrm{lsr}$) & $\req$ & Sum \tmb & Peak \tmb & Peak S/N & Parent ID \\
 &  &  & \degr & \degr & \kms & \kms & arcsec & K & K &  &  \\
\hline
G110.224+00.069 & G110 & 1 & 110.22371 & 0.06935 & -53.09 & 0.89 & 38.4 & 44592.5 & 20.8 & 45.6 & 1 \\
G110.122+00.087 & G110 & 2 & 110.12224 & 0.08703 & -51.03 & 1.08 & 49.8 & 127261.3 & 21.9 & 44.2 & 1 \\
G109.982-00.072 & G110 & 3 & 109.98246 & -0.0716 & -51.06 & 0.64 & 42.6 & 18207.2 & 16.2 & 37.9 & 1 \\
G110.194+00.012 & G110 & 4 & 110.19398 & 0.01226 & -50.41 & 1.58 & 40.3 & 44889.0 & 17.3 & 31.1 & 1 \\
G110.301+00.003 & G110 & 5 & 110.30127 & 0.00345 & -52.75 & 0.72 & 31.4 & 20078.0 & 14.1 & 29.9 & 1 \\
\hline
\end{tabular}
\label{tab:cat_hires}
\end{table*}

Table~\ref{tab:cat_hires} contains the first five rows of the \fwhires\ catalogue, with selected information given. We make full versions of the \fwlores, \fwhires, and \fwchimps\ catalogues available alongside this paper, and we detail the column descriptions in Appendix~\ref{app:table}. The format of the three catalogues is almost identical, with the exception that the \fwhires\ catalogue also lists the ID of the parent source in the \fwlores\ catalogue to allow the hierarchical information about the larger complexes to be retained.

\begin{figure*}
    % Figure_FellWalker.py
    \centering
    \vspace{-2mm}
    \includegraphics[width=0.93\textwidth]{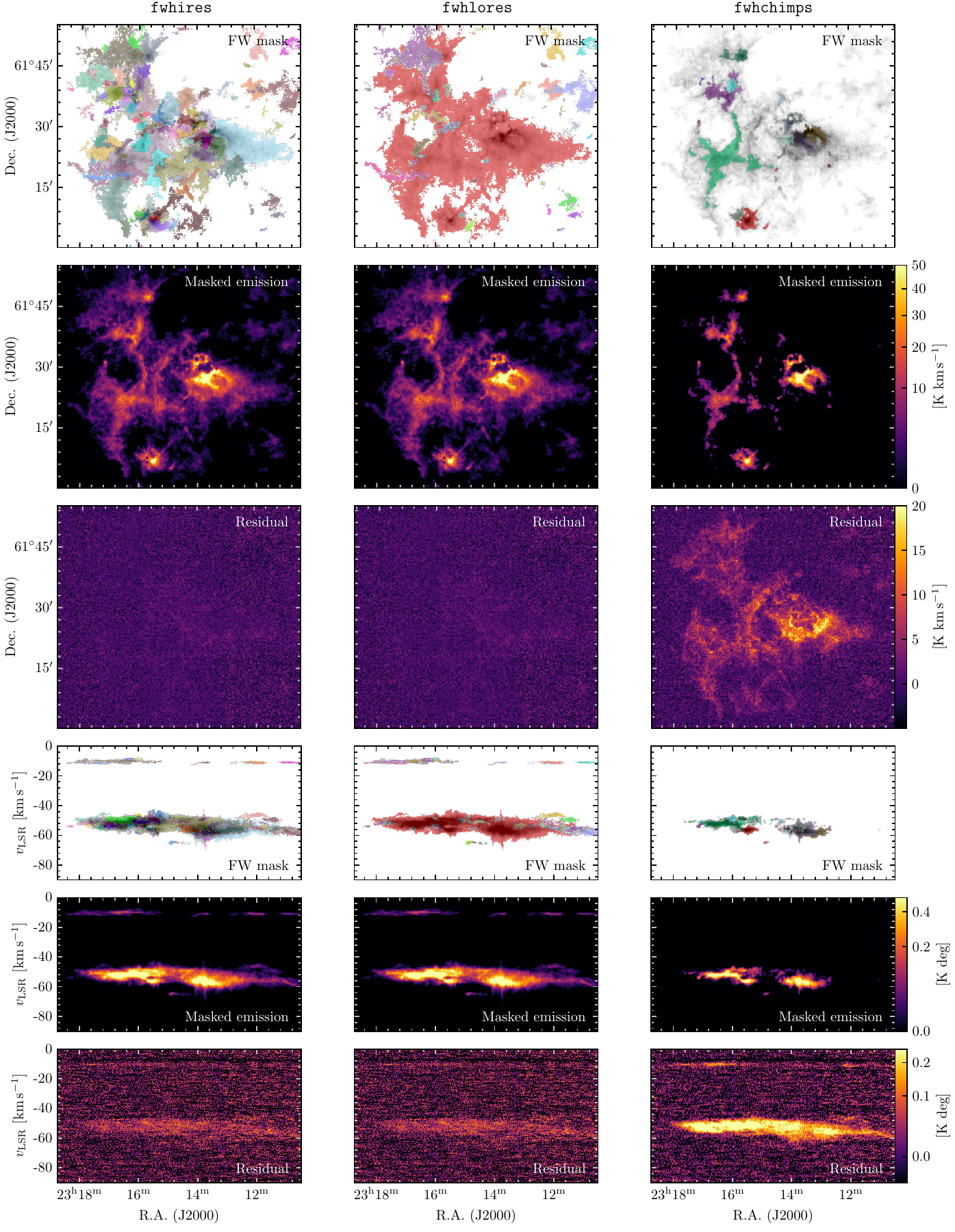}
    \caption{Each column of images shows different aspects of the \fwhires, \fwlores, and \fwchimps\ \fw\ source-extraction setups. \emph{Top row}: A greyscale representation of the \tco\ (3--2) velocity-integrated intensity map of NGC\,7538 overlaid with two-dimensional representations of the \fw\ masks. \emph{Second row}: Velocity-integrated intensity images from the \fw-masked cubes. \emph{Third row}: Residual velocity-integrated intensity map. \emph{Fourth row}: A greyscale representation of the \tco\ (3--2) declination-integrated position-velocity map overlaid with two-dimensional representations of the \fw\ masks. \emph{Fifth row}: Declination-integrated position-velocity maps of the masked cubes. \emph{Bottom row}: Residual declination-integrated position-velocity maps of the masked intensity. The images in each row are all on the same intensity scale, indicated by the colour bar on the right-most image.}
    \label{fig:fellwalker}
\end{figure*}

Fig.~\ref{fig:fellwalker} illustrates the differences between the source- extraction setups. The two-dimensional representations of the \fw\ masks clearly show the difference between the \fwhires\ and \fwlores\ extractions, illustrating that the same pixels of emission are recovered, but differ in their assignment to different structures. The \fwchimps\ extraction shows the substantial difference that the data quality makes in source extraction, and the necessity to have a like-for-like extraction in order to make meaningful comparisons to other data sets that take into account biases resulting from sensitivity. Both \fwhires\ and \fwlores\ recover much fainter emission than \fwchimps, which exists in the diffuse envelopes of the molecular clouds. The residual images give an idea of how complete the various extractions are. While \fwhires\ and \fwlores\ leave almost no visible residual in the integrated position-position intensity maps, those extractions still leave faint and unrecovered emission that is most clearly visible in the residual integrated position-velocity intensity maps.

\section{Results}
\label{sec:results}

\subsection{CO-to-H$_2$ conversion factors} \label{sec:XCO}

\begin{figure}
    \centering
    \includegraphics[width=\linewidth]{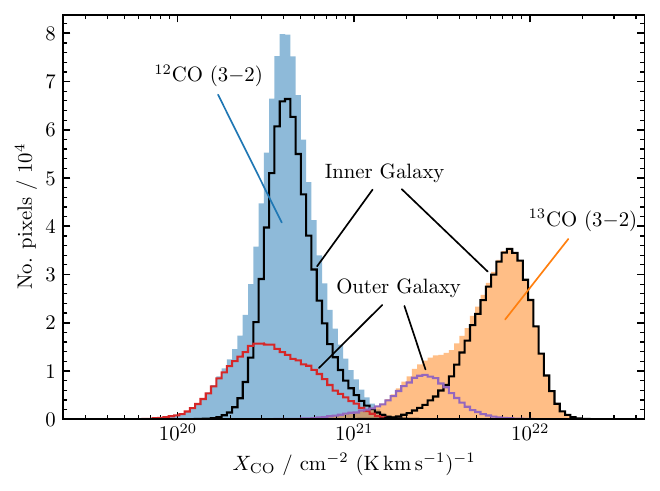}
    \caption{Histograms comparing the distributions of \xco\ for \twco\ and \tco\ (3--2) for the all pixels (shaded), and for the Inner and Outer Galaxy subsamples. The \tco\ (3--2) histogram has been scaled up by a factor of 2 for illustrative purposes.}
    \label{fig:XCOhist}
\end{figure}

\begin{figure*}
    \centering
    \includegraphics[width=\textwidth]{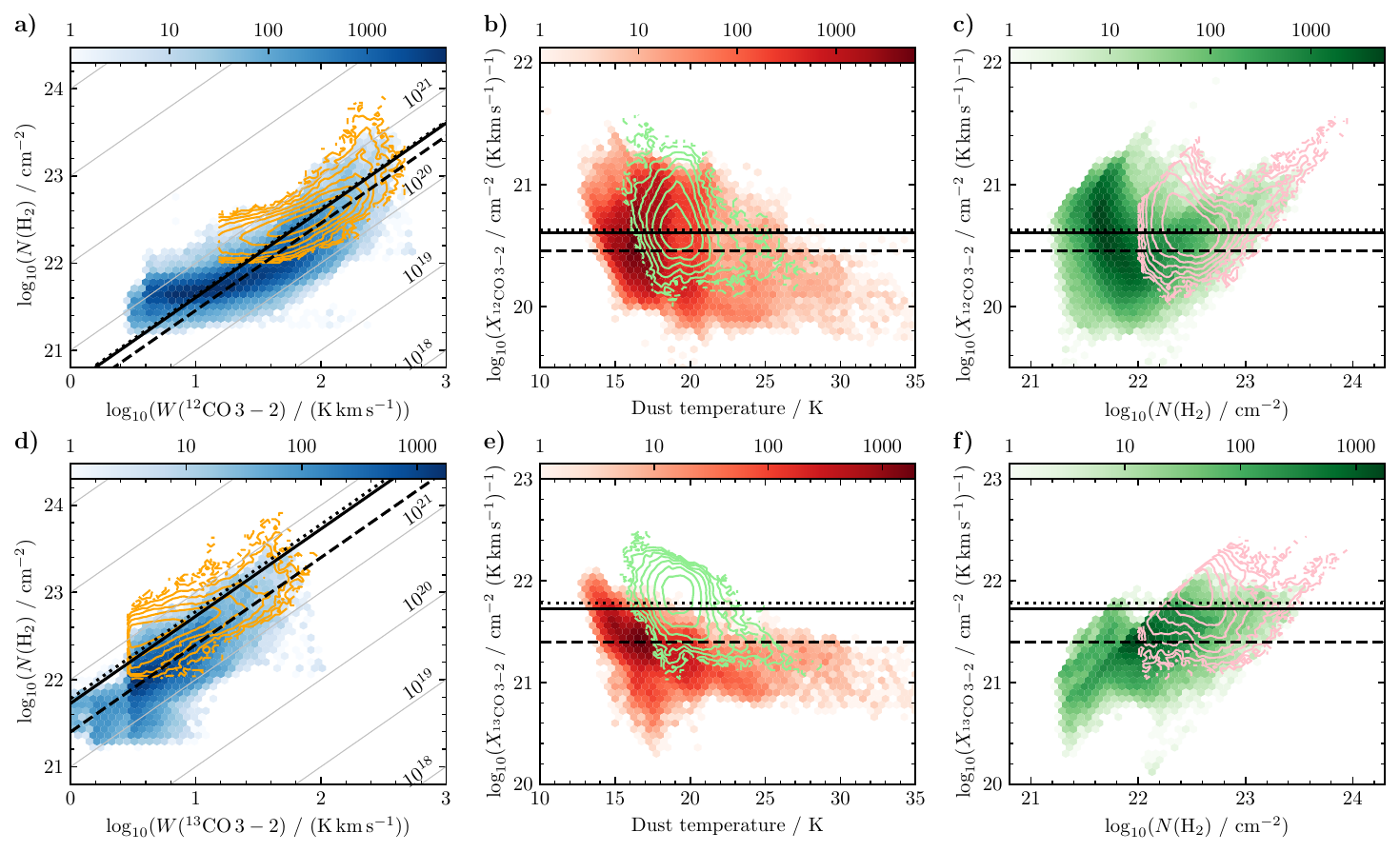}
    \caption{Two dimensional histograms comparing the pixel-by-pixel distributions of: (\emph{left column}) H$_2$ column density as a function of integrated intensity; (\emph{middle column}) \xco\ as a function of dust temperature; (\emph{right column}) \xco\ as a function of H$_2$ column density. The top and bottom rows examine the distributions for \twco\ and \tco\ (3--2) emission, respectively. In each case, the hexagonal histograms show the values for the Outer Galaxy regions, while the logarithmically spaced contours represent the Inner Galaxy region. The solid, dotted, and dashed lines in each panel show the global values (Method iv) from Section~\ref{sec:XCO}) of \xco\ derived for the combined, Inner Galaxy, and Outer Galaxy samples, respectively.}
    \label{fig:XCO}
\end{figure*}

The $X_\mathrm{CO}$ factor converts the integrated intensity of CO emission in a particular transition (and a particular isotopologue) to total molecular-hydrogen column density. In its general form:

\begin{equation} \label{eq:XCO}
    N(\mathrm{H}_2) = X_{\mathrm{CO}}\,W(\mathrm{CO})
\end{equation}

\noindent where $N(\mathrm{H}_2)$ is the molecular-hydrogen column density, and  $W(\mathrm{CO})$ is the integrated intensity of \twco\ (1--0) emission. For reference, \citet{Bolatto+13} recommend a typical value of $X_\mathrm{CO} = 2 \times 10^{20}$\,\xunits\ in the Milky Way disc. \xco\ condenses the wide range of environmentally varying excitation conditions that are likely to be found within a relatively large area of Galactic disc (typically used in external galaxies) into a single scaling relationship. Here, we explore the value of \xco\ for our \twco\ and \tco\ (3--2) emission lines (which we refer to as \xtwco\ and \xtco, respectively) within PAMS, and compare the results with those derived from similar data in an Inner Galaxy field. The Inner Galaxy region we used is centred at $\ell = 30$\degr -- which contains the massive star-forming complex W43 and cloud G29.96$-$0.02  -- using mosaics of the \twco\ COHRS \citep{Dempsey+13,Park+23} and \tco\ (3--2) CHIMPS data \citep{Rigby+16}. The latter were re-processed for the CHIMPS2 Inner Galaxy survey (Rigby et al. in preparation) on 6 arcsec pixels, with an effective resolution matching PAMS.

To calculate \xco\ values, we compared the moment 0 (velocity-integrated intensity) maps to maps of H$_2$ column density derived from greybody fitting with 160 and 250\,\mic\ imaging from the \emph{Herschel}/Hi-GAL survey following the method of \citet{Peretto+16}. The specific intensity $I$ at a frequency $\nu$ is related to H$_2$ column density via
\begin{equation}
    I_\nu = \mu_{\mathrm{H}_2} m_\mathrm{H} N(\mathrm{H}_2) \gamma^{-1} \kappa_0 \left(\frac{\nu}{\nu_0}\right)^{\beta} B_\nu(T_\mathrm{d}),
\end{equation}
where $\mu_{\mathrm{H}_2}$ is the mean molecular mass per H$_2$ (equal to 2.8 for molecular gas with a relative Helium abundance of 25 per cent), $m_\mathrm{H}$ is the mass of a hydrogen atom, $\gamma$ is the gas-to-dust mass ratio, $\kappa_\nu$ is the dust opacity at a reference frequency of $\nu_0$, $B_\nu(T_\mathrm{d})$ is the value of the Planck function evaluated at frequency $\nu$ for the dust temperature $T_\mathrm{d}$. First, maps of dust temperature are determined from the ratio of 160 to 250\,\mic\ flux densities after convolving the former to 18-arcsec resolution to match the latter:
\begin{equation}
    \frac{I_{\nu_{160}}}{I_{\nu_{250}}} = \left(\frac{250}{160}\right)^{\beta}  \left(\frac{B_{\nu_{160}}(T_\mathrm{d})}{B_{\nu_{250}}(T_\mathrm{d})}\right),
\end{equation}
for which we adopt $\beta=2$, and sample dust temperatures to the nearest 0.01\,K. We then generate maps of column density using the corresponding 250\,\mic\ map:
\begin{equation} \label{eq:coldens}
    N(\mathrm{H}_2) = \frac{I_{\nu_{250}} \gamma}{\mu_{\mathrm{H}_2} m_\mathrm{H} \kappa_0 \left(\frac{\nu}{\nu_0}\right)^{\beta} B_\nu(T_\mathrm{d})},
\end{equation}
using a value of $\kappa_0 = 12.0$\,cm$^2$\,g$^{-1}$ at 250\,\mic\footnote{The dust opacity is interpolated from Table 1, column 5 of \citet{Ossenkopf+Henning94}, but scaled down by a factor of 1.5, which was found by \citet{Kauffmann+10} to better reproduce extinction measurements.}, and a typical gas-to-dust mass ratio of $\gamma=100$.

The \twco\ and \tco\ cubes, and H$_2$ column-density maps were then smoothed to a common angular resolution of 20 arcsec, and the cubes were integrated over velocity ranges containing the emission to produce moment 0 (integrated intensity) maps. We masked the moment 0 maps below a contour level determined for each region in order to limit the impact of noise. The H$_2$ column-density maps were then resampled onto the same pixel grid as the moment 0 images so that the column densities and integrated intensities could be compared on a pixel-by-pixel basis. The overlap between Hi-GAL and PAMS meant that we could calculate $X_\mathrm{CO}$ factors for around 70 per cent of the survey area with S152, S157, and most of G110 falling out of the Hi-GAL coverage (see Fig.~\ref{fig:targets}).

We performed several calculations of \xtwco\ and \xtco\ for our regions using different methods; 

\begin{enumerate}
    \item We used equation (\ref{eq:XCO}), and performed a least-squares fit to the data points;
    \item We adapted the first approach to include a column-density offset (i.e. background correction):
    \begin{equation}
        N(\mathrm{H}_2) = X_{\mathrm{CO}}\,W(\mathrm{CO}) + N(\mathrm{H}_2)_\mathrm{bg}; 
    \end{equation}
    \item We calculated the median value of the individual \xco\ pixel values;
    \item We calculated a global average as:
    \begin{equation}
        X_\mathrm{CO} = \frac{\sum_{i}{N(\mathrm{H}_2)_i}} {\sum_{i}{W(\mathrm{CO})_i}}.
    \end{equation}
\end{enumerate}

\noindent We summarise these results in Tables~\ref{tab:X12CO} and \ref{tab:X13CO} for \twco\ (3--2) and \tco\ (3--2), respectively.

\begin{table*}
\renewcommand{\arraystretch}{1.2}
\centering
% X12CO_P16.tex
\caption{\xco\ values for \twco\ (3--2) in the various regions, with values listed for Methods i)--iv) outlined in Section~\ref{sec:XCO}. For Method ii) the background column-density value, $N(\mathrm{H}_2)$, is also listed. For Method iii) the uncertainties indicate the 16th--84th percentile range.}
\begin{tabular}{cccccc}
\hline
Region & i) \xtwco & ii) \xtwco & $N(\mathrm{H}_2)_0$ & iii) \xtwco & iv) \xtwco \\
 & \xunits & \xunits & \pcmm & \xunits & \xunits \\
\hline
G110 & $2.7 \times 10^{20}$ & $1.8 \times 10^{20}$ & $2.8 \times 10^{21}$ & $4.0^{+1.8}_{-1.4} \times 10^{20}$ & $3.4 \times 10^{20}$ \\
G135 & $1.4 \times 10^{20}$ & $5.4 \times 10^{19}$ & $2.8 \times 10^{21}$ & $2.8^{+3.9}_{-1.5} \times 10^{20}$ & $2.1 \times 10^{20}$ \\
IRAS\,02327 & $3.1 \times 10^{20}$ & $2.0 \times 10^{20}$ & $2.6 \times 10^{21}$ & $4.9^{+3.8}_{-2.1} \times 10^{20}$ & $4.0 \times 10^{20}$ \\
NGC\,7538 & $2.5 \times 10^{20}$ & $2.3 \times 10^{20}$ & $1.6 \times 10^{21}$ & $3.2^{+3.2}_{-1.3} \times 10^{20}$ & $2.8 \times 10^{20}$ \\
W3 & $2.2 \times 10^{20}$ & $1.5 \times 10^{20}$ & $4.0 \times 10^{21}$ & $3.4^{+1.3}_{-1.1} \times 10^{20}$ & $3.0 \times 10^{20}$ \\
W5 & $2.5 \times 10^{20}$ & $1.5 \times 10^{20}$ & $3.2 \times 10^{21}$ & $4.5^{+3.1}_{-2.3} \times 10^{20}$ & $3.5 \times 10^{20}$ \\
\hline
Inner Galaxy & $4.0 \times 10^{20}$ & $3.0 \times 10^{20}$ & $9.1 \times 10^{21}$ & $4.3^{+1.9}_{-1.0} \times 10^{20}$ & $4.2 \times 10^{20}$ \\
Outer Galaxy & $2.5 \times 10^{20}$ & $2.3 \times 10^{20}$ & $1.7 \times 10^{21}$ & $3.4^{+2.9}_{-1.4} \times 10^{20}$ & $2.9 \times 10^{20}$ \\
All & $3.8 \times 10^{20}$ & $3.4 \times 10^{20}$ & $3.7 \times 10^{21}$ & $4.1^{+2.1}_{-1.3} \times 10^{20}$ & $4.0 \times 10^{20}$ \\
\hline
\end{tabular}
\label{tab:X12CO}

\centering
% X13CO_P16.tex
\caption{\xco\ values for \tco\ (3--2) in the various regions, with values listed for Methods i)--iv) outlined in Section~\ref{sec:XCO}. For Method ii) the background column-density value, $N(\mathrm{H}_2)$, is also listed. For Method iii) the uncertainties indicate the 16th--84th percentile range.}
\begin{tabular}{cccccc}
\hline
Region & i) \xtco & ii) \xtco & $N(\mathrm{H}_2)_0$ & iii) \xtco & iv) \xtco \\
 & \xunits & \xunits & \pcmm & \xunits & \xunits \\
\hline
G110 & $2.2 \times 10^{21}$ & $1.5 \times 10^{21}$ & $5.3 \times 10^{21}$ & $2.3^{+1.3}_{-0.9} \times 10^{21}$ & $2.4 \times 10^{21}$ \\
G135 & $1.4 \times 10^{21}$ & $6.4 \times 10^{20}$ & $2.6 \times 10^{21}$ & $1.6^{+0.8}_{-0.5} \times 10^{21}$ & $1.6 \times 10^{21}$ \\
IRAS\,02327 & $2.5 \times 10^{21}$ & $1.9 \times 10^{21}$ & $2.4 \times 10^{21}$ & $3.2^{+1.2}_{-1.0} \times 10^{21}$ & $2.9 \times 10^{21}$ \\
NGC\,7538 & $2.4 \times 10^{21}$ & $2.2 \times 10^{21}$ & $2.2 \times 10^{21}$ & $2.5^{+1.1}_{-0.9} \times 10^{21}$ & $2.5 \times 10^{21}$ \\
W3 & $1.2 \times 10^{21}$ & $7.8 \times 10^{20}$ & $8.1 \times 10^{21}$ & $2.2^{+1.2}_{-1.0} \times 10^{21}$ & $2.0 \times 10^{21}$ \\
W5 & $1.8 \times 10^{21}$ & $1.5 \times 10^{21}$ & $2.3 \times 10^{21}$ & $2.0^{+1.2}_{-0.8} \times 10^{21}$ & $2.0 \times 10^{21}$ \\
\hline
Inner Galaxy & $4.8 \times 10^{21}$ & $3.0 \times 10^{21}$ & $2.1 \times 10^{22}$ & $7.0^{+2.9}_{-2.5} \times 10^{21}$ & $6.0 \times 10^{21}$ \\
Outer Galaxy & $2.4 \times 10^{21}$ & $2.2 \times 10^{21}$ & $1.8 \times 10^{21}$ & $2.4^{+1.1}_{-0.9} \times 10^{21}$ & $2.5 \times 10^{21}$ \\
All & $4.3 \times 10^{21}$ & $2.9 \times 10^{21}$ & $1.7 \times 10^{22}$ & $6.1^{+3.3}_{-3.3} \times 10^{21}$ & $5.3 \times 10^{21}$ \\
\hline
\end{tabular}
\label{tab:X13CO}
\end{table*}

We show the distributions of \xco\ pixel values across PAMS in Fig.~\ref{fig:XCOhist}, in which it is apparent that the Outer Galaxy values are lower than the Inner Galaxy values for both \twco\ and \tco\ (3--2), with the discrepancy being larger for the latter. Fig.~\ref{fig:XCO} illustrates several additional aspects of these results by exploring the correlation between individual \xco\ pixel values and other properties. In Panels a) and d), we show the distributions of pixel values of H$_2$ column density as functions of the integrated intensity of \twco\ and \tco\ (3--2) in the corresponding pixel. In both cases, it is clear that the distributions are not completely linear (i.e. simple power-laws in log-space) and there is considerable scatter. We see a flattening of the distributions at low integrated intensity that is more prominent in \twco, and indicative of the column-density background that is detected in the spaced-based \emph{Herschel} observations, but not discernible in the ground-based CO observations due to their sensitivity. At the bright end of the distributions, the column density also curves upwards -- and this behaviour is stronger in \twco\ (3--2) -- indicating a saturation of CO emission as a consequence of the emission becoming optically thick at the highest column densities. The global \xco\ values (Method iv) are plotted as black lines, and these are the values that are representative of the kind of \xco\ values that are used in studies of extragalactic systems, where star-forming complexes may be unresolved. The scatter around the global values is clearly not random, and we explore the origin by examining the \xco\ distributions as functions of dust temperature and column density in Panels b), c), e), and f). 

The distributions of individual \xco\ values are very broad, spanning between one and two orders of magnitude in both isotopologues. For all cases, the distributions of dust temperature are also broad for a given \xco\ value. One very noticeable trend is that the lower envelope of the dust temperature distribution is $\sim$5\,K higher in the Inner Galaxy samples (contours). It is also of particular note that the peak of the distribution of \xtco\ values in the Inner Galaxy is a factor of $\sim$2--3 higher than the global \xtco\ value for the Outer Galaxy, suggesting that global values of \xco\ are likely to be weighted towards particular ISM conditions that are conducive to bright CO emission, but which are not representative of typical conditions by area.

For \xtwco, the range of values returned from our different methods are fairly comparable, ranging between 2.3--4.2$\times 10^{20}$\xunits. Each of the methods systematically returns a higher value of \xtwco\ for the Inner Galaxy than for the Outer Galaxy region although the distributions overlap substantially (see Fig.~\ref{fig:XCOhist}). Given the different calculation methods, we suggest that -- in principle -- our Method i) values are most appropriate for resolved studies, and Method iv) values are more appropriate for unresolved studies. Given the insignificant difference for \twco\ (3--2), we recommend the usage of \xtwco\,=\,\xtwcounresolved\ for both resolved studies and unresolved studies, noting that range of values varies by around $\pm50$ per cent (as per Method iii). This is in excellent agreement with \citet{Colombo+19} who derived by their value by applying an average 3--2/1--0 line-ratio measurement to the \citet{Bolatto+13} \xco\ value listed above using the wider COHRS data set of which we take a subset for our Inner Galaxy sample.

The picture is more varied for \xtco, with values between 2.2--7.0$\,\times 10^{21}$\xunits\ across the different regions and calculation methods. All four methods return higher values in the Inner Galaxy than in the Outer Galaxy by a factor of $\sim$2--3. The higher column-density background in the Inner Galaxy, which arises as a consequence of the greater number of spiral arms present along the line of sight and a greater column of Galactic disc, appears to be the primary driver of the different recovered values. This variation is clearly illustrated in Fig.~\ref{fig:XCO}d where the global average lines (for Method iv) are more widely separated for the Inner and Outer Galaxy. The peak of the \xtco\ distributions in Figs.~\ref{fig:XCOhist}, \ref{fig:XCO}e, and \ref{fig:XCO}f between the Inner and Outer Galaxy are noticeably different. In the Inner Galaxy, it appears that the higher dust-temperature background of $\sim$20\,K compared with $\sim$17\,K in the Outer Galaxy, along with a higher background of column density contribute to the higher \xtco\ values. The single dust temperature is less likely to represent the underlying gas conditions at this Inner Galaxy position, where a greater fraction of the Galactic disc is located within the column. We note that gas and dust temperatures are only likely to be coupled at densities greater than $\sim n(\mathrm{H}_2) > 10^{4.5}$\pcmmm\ \citep{Goldsmith01}, so that for much of the gas in molecular clouds -- especially their envelopes -- the CO excitation temperature is unlikely to follow the dust temperature. We will explore CO excitation temperatures, along with a local thermodynamic equilibrium (LTE)-based \xco\ derivation in a future paper. 

For resolved studies, we recommend the usage of our Method i) value of \xtcoresolved\ as a representative value for \xtco, and for unresolved studies, we recommend a value of \xtcounresolved. As for \twco, these \xco\ factors should always be used with the knowledge that their derivation is biased towards (relatively) hot high-column-density gas where CO is brightest on large scales, but will not accurately account for column densities in other environments. It is also important to recognise that a Galactic gradient of \xco\ values is evident in our data, which we will discuss further in Section~\ref{sec:discussion}. In all cases, we recommend that a multiplicative factor of 1.5 be adopted for the uncertainty in the \xco\ values to encapsulate the 1-$\sigma$ spread of the values.

\subsection{Molecular-cloud properties} \label{sec:properties}

In this Section, we explore some of the basic properties of molecular-cloud structures within PAMS, and compare those with an Inner Galaxy reference from the CHIMPS survey \citep{Rigby+19}. The reference sample was restricted to include only those clouds with distances between 2--4\,kpc which approximately matches the range in distance of the PAMS sources and thus limits the effect of distance biases. Since CHIMPS covers a longitude range of $28\degr \lesssim \ell \lesssim 46\degr$, the distance limitation results in the Inner Galaxy sample covering a range in Galactocentric radius of $5 \lesssim \rgc \lesssim 7$\,kpc, compared with roughly 9--10\,kpc for the PAMS Outer Galaxy sample. We also applied a minimum peak SNR criterion to all of our catalogues in order to emulate the `reliability' flagging that was made in the CHIMPS catalogue. In the CHIMPS catalogue, 95 per cent of sources with the highest-reliability flags have a peak SNR greater than 9 and, similarly, 95 per cent of sources with the lowest reliability flag have a peak SNR less than 9, and so we adopt this value for our cut. This conservative cut helps eliminate potentially spurious sources that \fw\ can produce at low SNR, which often appear as separated islands of low-intensity emission (which we refer to as `archipelagos'). We adopt the same cut for the CHIMPS Inner Galaxy sample (as opposed to using the reliability flags directly) for consistency. By applying these cuts, the \fwhires, \fwlores, \fwchimps, and Inner Galaxy catalogues were reduced from 948, 483, 231, and 4999 sources to 499, 80, 140, and 865 sources, respectively.

\subsubsection{Source radii} \label{sec:radii}

\begin{figure*}
    % Analysis_scaling.py
    \centering
    \includegraphics[width=\textwidth]{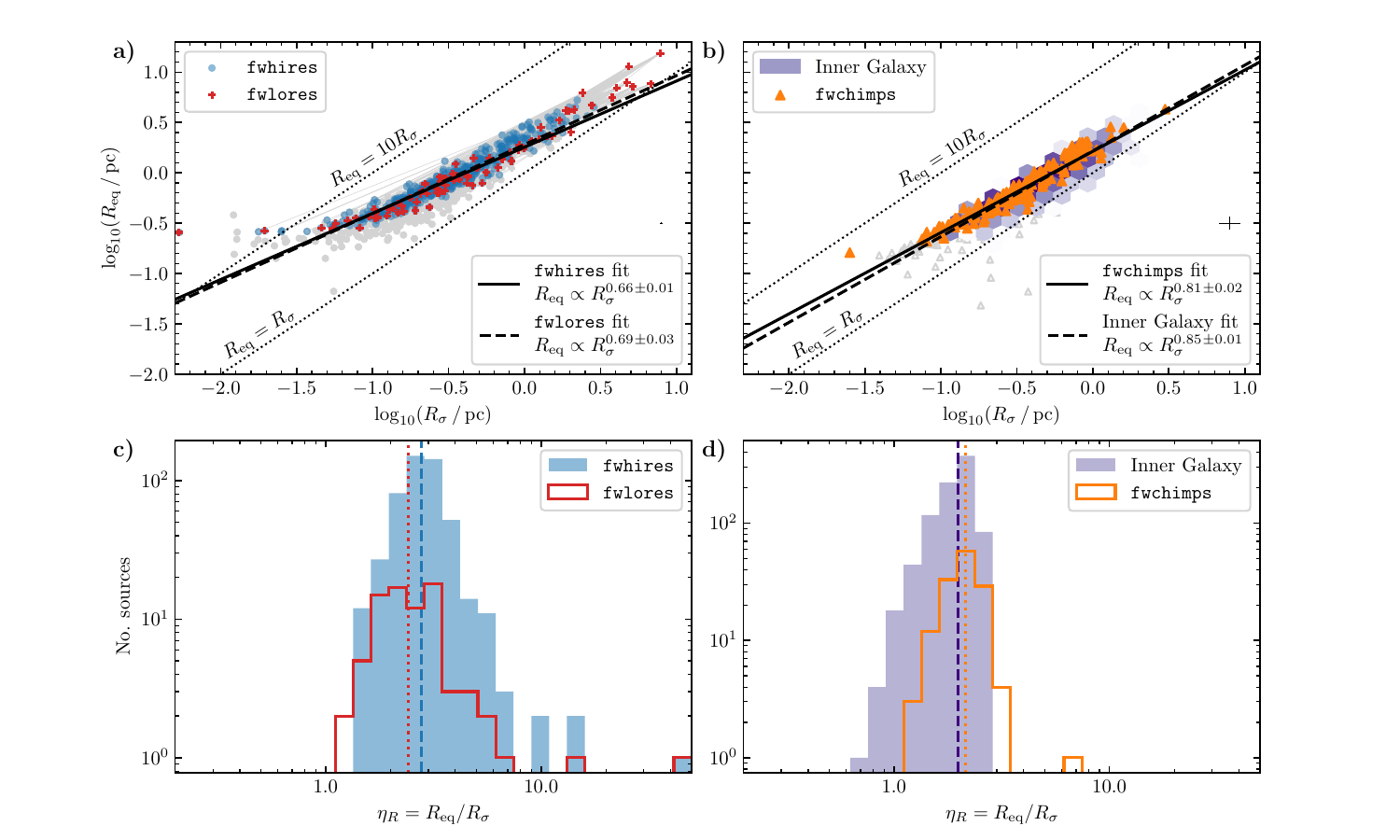}
    \caption{Comparison of intensity-weighted rms radii $\rsig$ and equivalent radii $\req$ for sources extracted from PAMS and the $2 \ge d \leq 4$\,kpc Inner Galaxy sample for CHIMPS. \textbf{a)} Comparison between radii derived from the \fwhires\ and \fwlores\ \fw\ extractions. Where a \fwhires\ source is a fragment of a larger source in the \fwlores\ extraction, it is linked to its parent with a grey line. \textbf{b)} Comparison of radii from the \fwchimps\ \fw\ PAMS source extraction, (orange triangles), and the CHIMPS Inner Galaxy sample (hexagonal histogram). The colours of the hexagonal histogram data are have been normalised with a logarithmic intensity scaling. In Panels a) and b) the solid and dashed lines show the best fitting power-law relationships, and we show the points not used in the fit (with peak SNR $<9 $) as empty grey symbols. Median error bars are indicated to the right. Panels \textbf{c)} and \textbf{d)} show the distributions of $\etarad$ for the points used in the fits in Panels a) and b), respectively, with median values indicated by dashed and dotted vertical lines for the filled and empty histograms, respectively.}
    \label{fig:radius}
\end{figure*}

There are several ways of reporting the size of molecular clouds, which each have drawbacks due to the intrinsic difficulty of representing the complex morphologies and intensity distributions with simple metrics that can be given in a catalogue. These differences may have implications for the way in which key scaling relationships, such as the size--linewidth relationship \citep{Larson81}, are compared between different data sets and so we explore two of them briefly here.

The intensity-weighted radius, $\rsig$, is given by:
\begin{equation}
    \rsig = d \, \sqrt{\sigma_x \sigma_y},
\end{equation}
\noindent where $d$ is the source distance and $\sigma_x$ and $\sigma_y$ are the intensity-weighted rms dispersions in the $x$- and $y$-axes of the image (in the case of PAMS, R.A. and Dec., respectively), deconvolved by the effective beam size. In the case of a perfect Gaussian source detected at a high signal-to-noise ratio, $\rsig$ would give the standard deviation of the source profile, equivalent to the FWHM / $\sqrt{8 \ln{2}}$. An alternative is the radius of a circle with the equivalent angular area, $A$, of the source. The equivalent radius is:
\begin{equation}
    \req = d \, \sqrt{A / \pi}.
\end{equation}
\noindent Again, the angular radius must first be deconvolved by the effective beam size before scaling to the relevant distance. We also define 
\begin{equation}
    \etarad = \req / \rsig
\end{equation}
to capture the ratio of the two measurements. These two radii differ in approach because $\rsig$ depends upon the intensity distribution of the source, while $\req$ depends only upon the area of the footprint of the source. A cloud with a compact and bright centre surrounded by diffuse emission will, therefore, have a much smaller value of $\rsig$ than $\req$. $\req$ is more easily impacted by the sensitivity of the observations, and will recover larger values in deeper observations that detect more diffuse emission. By contrast $\rsig$ varies less across observations of different sensitivity and so we generally favour this prescription (and indeed $\etarad$ is weakly correlated with peak SNR). Making both measurements of the radius will allow maximum compatibility with other measurements in the literature which use either method, and environmental trends may also reveal themselves in the relationship between these two measurements.

In Fig.~\ref{fig:radius} we show the relationship between $\rsig$ and $\req$ for our various \fw\ source extractions along with their $\etarad$ distributions, and with a comparison to the values reported by \citet{Rigby+19} for the Inner Galaxy from the distance-limited CHIMPS sample. The much larger CHIMPS sample is illustrated as a two-dimensional hexagonal histogram to allow the point density to be seen more easily. In all cases, we find that the relationship between $\rsig$ and $\req$ is well fitted by a power law. We performed a power-law fit to the two radii types for each sample using an orthogonal distance regression\footnote{Using \href{https://docs.scipy.org/doc/scipy/reference/odr.html}{\texttt{scipy.odr}}.} to account for the uncertainties on both variables. The fit results are reported in Table~\ref{tab:fitresults}. We find that the \fwhires\ and \fwlores\ source extractions produce essentially identical relationships, with power-law indices of $0.66\pm0.01$ and $0.69\pm0.03$, respectively. This is unsurprising because the two extractions are very similar, differing only in the level of fragmentation that they allow (\fwhires\ structures exist within \fwlores\ structures).

\begin{table*}
\centering
\caption{The results of fitting to the relationships described in Sections~\ref{sec:radii} and \ref{sec:scaling}. The fits were made in log space, where $\log_{10}(y) = m \log_{10}(x) + \log_{10}(k)$ for a relationship $y = k x^m$. Uncertainties from the fitting algorithm on $m$ and $\log_{10}(k)$ are also provided.}
\begin{tabular}{ccccccc}
\hline
Sample & $x$ & $y$ & $\log_{10}(k)$ & $\Delta \log_{10}(k)$ & $m$ & $\Delta m$ \\
\hline
\fwhires & $\rsig$ / pc & $\req$ / pc & 0.255 & 0.007 & 0.657 & 0.010 \\
\fwlores & $\rsig$ / pc & $\req$ / pc & 0.281 & 0.019 & 0.685 & 0.025 \\
\fwchimps & $\rsig$ / pc & $\req$ / pc & 0.214 & 0.012 & 0.807 & 0.020 \\
Inner Galaxy & $\rsig$ / pc & $\req$ / pc & 0.218 & 0.005 & 0.852 & 0.011 \\
\hline
\fwhires & $\rsig$ / pc & $\Delta v$ / \kms & 0.080 & 0.016 & 0.284 & 0.027 \\
\fwlores & $\rsig$ / pc & $\Delta v$ / \kms & 0.040 & 0.020 & 0.405 & 0.033 \\
\fwchimps & $\rsig$ / pc & $\Delta v$ / \kms & 0.121 & 0.026 & 0.437 & 0.052 \\
Inner Galaxy & $\rsig$ / pc & $\Delta v$ / \kms & 0.142 & 0.010 & 0.478 & 0.024 \\
\hline
\fwhires & $\rsig$ / pc & $M$ / \msun & 3.558 & 0.036 & 2.010 & 0.057 \\
\fwlores & $\rsig$ / pc & $M$ / \msun & 3.252 & 0.055 & 1.779 & 0.081 \\
\fwchimps & $\rsig$ / pc & $M$ / \msun & 3.824 & 0.055 & 2.146 & 0.096 \\
Inner Galaxy & $\rsig$ / pc & $M$ / \msun & 3.511 & 0.029 & 2.490 & 0.067 \\

\hline
\end{tabular}
\label{tab:fitresults}
\end{table*}

The relationship between the two radius types for the \fwchimps\ and the Inner Galaxy samples are also non-linear and with power-law indices of $0.81\pm0.03$ and $0.85\pm0.01$, respectively, and are consistent with each other. This suggests that there are no significant differences in the mean emission profile for sources at Galactocentric distances of 9--10\,kpc compared with sources at 5--7\,kpc.

The difference between the fit to the \fwchimps\ data and the \fwlores\ and \fwhires\ is caused by the differences in data quality and to the \fw\ parameter setup, which we have modified in PAMS compared with CHIMPS. We note that some of the non-fitted points in Figs.~\ref{fig:radius}a and b  show unusually high or low values of $\etarad$. In cases where $\rsig$ is much larger than $\req$, these are the archipelago sources that \fw\ identifies at low SNR. The sources with much larger $\req$ than $\rsig$ look like well-recovered sources that are diffuse and have flat emission profiles, and these are over-represented in the data due to a selection bias in \fw; the requirement for sources to have a minimum number of pixels above the intensity defined as the noise level prefers diffuse over compact sources. Such sources are also more likely to be the beneficiary of flux boosting effects, in which positive contributions to the emission from the noise may be represented in the data, but negative contributions will not.

\citet{Rigby+19} reported a median value of $\etarad = 2.0$ across the full CHIMPS sample, and the same value for our distance-limited Inner Galaxy sample (Fig.~\ref{fig:radius}d). For \fwchimps, the figure is slightly larger at 2.1. By contrast, the \fwhires\ and \fwlores\ median $\etarad$ values are larger at 2.8 and 2.4, respectively, which is expected for more-sensitive data. Overall, the emission profiles of the PAMS sources are similar to those of the Inner Galaxy sample, suggesting that any differences in emission characteristics between 5--7 and 9--10\,kpc are mild.

\subsubsection{Scaling relationships} \label{sec:scaling}

\begin{figure*}
    \centering
    % Analysis_scaling.py
    \includegraphics[width=\textwidth]{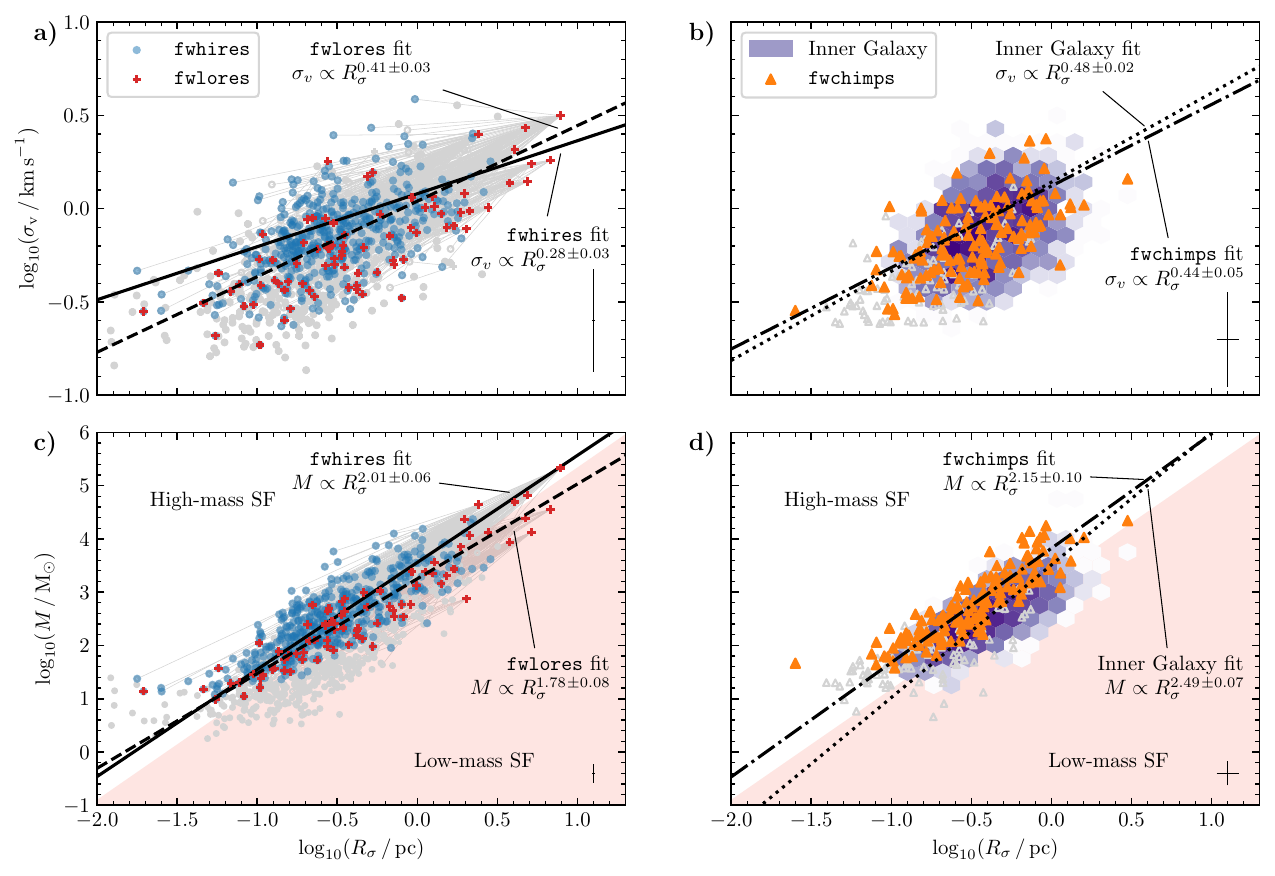}
    \caption{\emph{Top row}: Size--linewidth relationships for PAMS and CHIMPS sources. \emph{Bottom row}: mass--radius relationships for PAMS and CHIMPS clumps. For both rows, the figures in the left column show the distribution of points from the \fwhires\ and \fwlores\ extractions, along with their lines of best fit. The figures in the right column show the distribution of sources from CHIMPS as hexagonal two-dimensional histograms, and the \fwchimps\ PAMS extraction as orange triangles. The colours of the hexagonal histogram data are have been normalised with a logarithmic intensity scaling. The shaded area denotes the region that has been empirically found to be devoid of high-mass star formation in local star-forming regions \citep[adapted from][]{Kauffmann+Pillai10}. Median error bars are shown to the lower right in each panel. Points not used in the fitting are shown in grey.}
    \label{fig:scaling}
\end{figure*}

Fig.~\ref{fig:scaling} shows the size--linewidth and mass--radius relationships for different samples of our PAMS data, and with a comparison to the same distance-limited Inner Galaxy sample as in the previous section. We also explored the differences in cloud properties that might arise from our different observations and \fw\ parameter settings by showing the relationships for each of the \fwhires, \fwlores, and \fwchimps\ extractions. Masses were derived according to:
\begin{equation}
    M = \mu_{\mathrm{H}_2} m_\mathrm{H} d^2 X_\mathrm{^{13}CO\,3-2} \int W (^{13}\mathrm{CO}\,3-2)\, \mathrm{d}\Omega, 
\end{equation}
\noindent where $\mu_{\mathrm{H}_2}$ is the molecular weight per hydrogen molecule, with a value of 2.8 (accounting for a 71 per cent abundance of hydrogen, 27 per cent helium, and 2 per cent metals), $m_\mathrm{H}$ is the mass of a hydrogen atom, $d$ is the source distance, $X_\mathrm{^{13}CO\,3-2}$ is the $X$-factor calculated for \tco\ (3--2) derived in Section~\ref{sec:XCO}, $W (^{13}\mathrm{CO}\,3-2)$ is the integrated intensity of \tco\ (3--2) per pixel, and d$\Omega$ is the angular area of a pixel. We adopted the Method i) values of \xtco\ = $4.8\times10^{21}$ and $2.4\times10^{21}$\xunits\ as our best overall estimate for resolved regions in the Inner and Outer Galaxy samples (i.e. CHIMPS and PAMS), respectively, with a factor of 1.5 uncertainty. We note that although \citet{Rigby+19} calculated clump masses for the CHIMPS sources using an LTE analysis, we apply the same \xco\ derivation here for consistency with the PAMS data. We will calculate the masses of PAMS sources using LTE analysis for a comparison with the CHIMPS LTE masses and excitation conditions in a future paper.

We fitted power laws to the size--linewidth and mass--radius relationships using orthogonal distance regression, as in Section~\ref{sec:radii}, and present the results in Table~\ref{tab:fitresults}. Fig.~\ref{fig:scaling}a shows that the size--linewidth relationships\footnote{For the uncertainty on the linewidth, we use the difference between the measured linewidth and the \textit{corrected linewidth} that depends on the peak SNR of the source, which is given by $\sigma_\mathrm{v}^* = \sigma_\mathrm{v} (13.3 / (\mathrm{SNR} + 5.5)$. This empirical relationship was reported in Eq. 8 of \citet{Rigby+19}, and accounts for the clipping of the linewidth caused by the detection thresholds used by \fw\ (and any other source-extraction algorithms that do not explicitly model emission profiles, e.g. dendrograms), though the correction factor was erroneously described as the uncertainty itself.} for the \fwlores\ and \fwhires\ extractions show similar relationships, considering the uncertainties. Although \fwlores\ has a slightly steeper relationship, for the most part, the \fwhires\ sources have higher linewidths for a given size. This is a result of the greater level of fragmentation allowed in the \fwhires\ extraction, where larger linewidths are recovered due to intra-cloud variations that are averaged out in the corresponding \fwlores\ extraction. This is illustrated by the lines connecting the \fwhires\ sources to their \fwlores\ parent. For example, the NGC\,7538 region catalogues contain 58 and 132 entries in the \fwlores\ and \fwhires\ extractions, respectively, but the bulk of the mass (95 per cent of the \tco\ emission) is recovered as a single source in the \fwlores\ catalogue, which is the parent structure of 73 of the sources (i.e. around half) featuring in the \fwhires\ catalogue. The size--linewidth relationships for the \fwchimps\ and CHIMPS Inner Galaxy samples are almost identical.

The mass--radius relationships show more significant differences. Comparing the \fwlores\ and \fwhires\ source extractions, the \fwhires\ sources have a steeper power-law index of 2.01 compared with 1.78 to \fwlores, but again the difference is not significantly different when considering the uncertainties. The weak differences in the distributions correspond to a scale dependence in the density distributions of the sources; we note the greater frequency of clump-scale substructures -- with sizes of $\lesssim 3$\,pc and masses of $\lesssim 10,000$\,\msun -- in the \fwhires\ distribution. When comparing the \fwchimps\ and Inner Galaxy samples, the power-law index is steeper in the Inner Galaxy compared with the Outer Galaxy, with a value of 2.49 for CHIMPS compared with 2.16 for PAMS, but they are consistent within the uncertainties. This consistency indicates that the mean density profile of sources in the two samples are similar. 

In Fig.~\ref{fig:scaling}, we also illustrate the relationship of \citep{Kauffmann+Pillai10} that delineates the parameter space that is empirically found to be devoid of high-mass star formation (HMSF) in nearby clouds. The original relationship was devised in terms of $\req$, with $m(r) \leq 870\,M_\odot (\req/\mathrm{pc})^{1.33}$ as the limit. Since we have favoured $\rsig$, the relationship requires modification for application to our data, and we adopt $\req = 1.75 \rsig^{0.75}$ -- as the average of the fitted parameters in Section~\ref{sec:radii} -- and our adapted relationship is therefore $m(r) \leq 1830\,M_\odot (\req / \mathrm{pc})^{2.08}$. The majority of sources in the samples fitted in Fig.~\ref{fig:scaling} are capable of forming high-mass stars, with 79 and 61 per cent for the fitted \fwhires\ and \fwlores\ samples, respectively. The proportion appears to be higher, in fact, than is the case for the CHIMPS survey, with 96 and 63 per cent of the sources from the fitted CHIMPS and \fwchimps\ samples, respectively, satisfying the condition. We will explore why this might be the case in Section~\ref{sec:discussion} for HMSF.

\subsection{Galactocentric dependence} \label{sec:galactocentric}

The combination of PAMS and CHIMPS allows for an expanded study of properties as a function of Galactocentric distance. The longitude coverage of CHIMPS means that the only clouds in the survey that lie outside the solar circle are at the far side of the Galaxy, with distances in the range $\sim$12--17\,kpc. Consequently, those clouds are both few in number, and sample only the extreme high-column-density (and therefore high-mass) end of the underlying distributions due to Malmquist bias. The PAMS data, therefore, make an important contribution to Galactocentric trends by significantly improving the population statistics at $D_\mathrm{GC} \sim$9--10\,kpc, with much improved spatial resolution.

\begin{figure*}
    \centering
    % Analysis_Galactocentric.py
    \includegraphics[width=\linewidth]{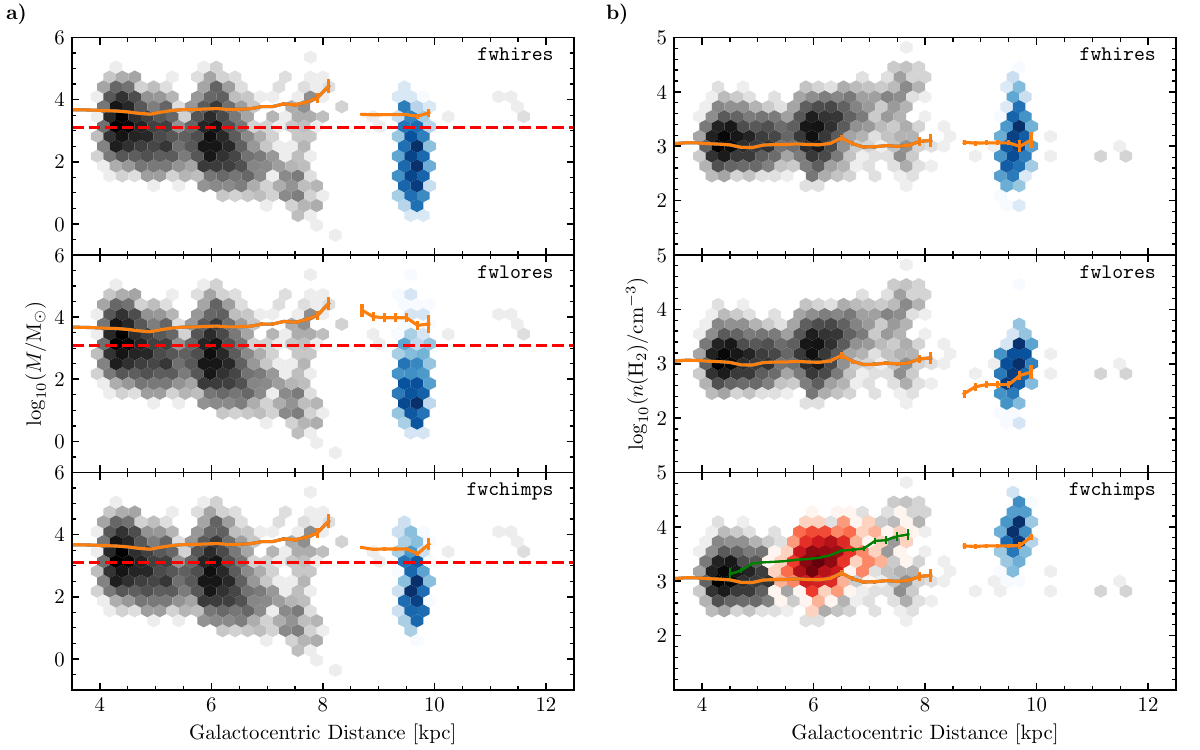}
    \caption{Logarithm of mass (\emph{left column}) and density (\emph{right column}) as a function of Galactocentric distance for clouds within CHIMPS \citep[greyscale histogram][]{Rigby+19}, and PAMS (blue histogram; this paper), for the \fwhires, \fwlores, and \fwchimps\ extractions from PAMS, respectively. The red dashed line indicates the sample mass-completeness limited, adapted from the CHIMPS calculation. The solid orange lines indicate the moving average in 0.2-kpc-wide bins for clouds that satisfy the mass-completeness limit. The lower-right panel also highlights CHIMPS data in the distance range 2--4\,kpc, with a secondary radial trend indicated by the thin green line.}
    \label{fig:Galactocentric}
\end{figure*}

In Fig.~\ref{fig:Galactocentric}a we display the mass distributions from CHIMPS (greyscale hexagonal histogram) and PAMS (blue hexagonal histogram). For both PAMS and CHIMPS, we take the masses calculated from our \xtco\ factor from Section~\ref{sec:XCO}. We determined mean radial trends on a subsample of the data by first excluding all sources with peak SNRs of less than 9, as in Sections~\ref{sec:radii} and \ref{sec:scaling}, and applying the mass-completeness threshold derived by \citet[][their equation B.1]{Rigby+19} -- shown as the red dashed line -- removing all sources at $d > 12$\,kpc, and all sources with $M < 1250$\,M$_\odot$. The completeness threshold should remove some of the effects of Malmquist bias. The solid orange line in Fig.~\ref{fig:Galactocentric} shows the moving average value of $\log_{10}(M/\mathrm{M}_\odot)$, with a window size of 0.1\,kpc, to all sources in the subsample. It is apparent that the PAMS clumps are  consistent with their Inner Galaxy counterparts, and we do not see any systematic trends over the range of $\rgc$ probed. This is not affected by the different \fw\ parameter configurations. As far as we can tell with these data, the cloud-mass distributions in \tco\ (3--2) do not vary with Galactocentric distance.

In Section~\ref{sec:scaling}, we found slight differences in the power- law index of the mass--radius relationship between PAMS and CHIMPS clouds, indicating that the two populations may have slightly different density profiles. We therefore examine the distribution of average volume densities in Fig.~\ref{fig:Galactocentric}b, calculated by:
\begin{equation} \label{eq:density}
    n(\mathrm{H}_2) = \frac{3}{4 \pi} \frac{M}{\mu_{\mathrm{H}_2} m_\mathrm{H} \req^3}.
\end{equation}
\noindent Here we adopt the equivalent radius to ensure that the mass and radius are calculated for the same volume. We calculated average Galactic radial profiles using the same reduced sample above containing only sources satisfying the minimum mass and SNR criteria, which are, again, shown in orange. The distribution of PAMS densities for the \fwhires\ extraction is consistent with the Inner Galaxy, whereas for \fwlores\ the mean values are a factor of $\sim$2 lower. The latter extraction suppresses small-scale structure, and so reports cloud complexes that evidently have lower mean densities (when using $\req$ for the calculation) due to the extended low-column-density envelopes that PAMS is more sensitive to. We therefore caution here that comparing distributions such as these derived from different source-extraction methods will almost certainly result in differences. When we compare like-for-like extractions, the PAMS densities within the \fwchimps\ distribution are a factor of $\sim$4 higher than their CHIMPS counterparts, which is somewhat surprising, and we see the same offset when considering the binned radial trend. This is explainable as a distance-related bias: In the CHIMPS sample, the relationship between Galactocentric and heliocentric distances bifurcates \citep[see Fig. 12g of][]{Rigby+19} due to the overlapping of sources in the near and far sides of the Galactic disc in velocity. This bifurcation is visible in the distribution of mean densities in Fig.~\ref{fig:Galactocentric}b at $\rgc \approx 8$\,kpc, with sources at the far distance occupying the lower-density branch and vice versa for sources at the near distance. This explains this apparent trend. In the lower-right panel we demonstrate this by plotting a secondary radial trend where we limited the CHIMPS sample to sources within the same 2--4\,kpc distance range that we have in our PAMS sample, and it is clear that the radial trend is consistent with that of the \fwchimps\ extraction of PAMS. The mean densities are particularly sensitive to distance effects when using a clump-scale extraction, with more internal substructures being discernible in closer sources. Extraction techniques which prefer to recover the largest scale structures such as \fwlores, or {\sc scimes} \citep[e.g][]{Colombo+15,Colombo+19,Duarte-Cabral+21,Rani+22} are therefore more suitable for examining Galactocentric gradients.

\section{Discussion} \label{sec:discussion}

% Sun+24 for MWISP Galactocentric trends
\subsection{Galactocentric gradients in \xco}

In Section~\ref{sec:XCO}, we calculated \xco\ factors for \twco\ and \tco\ (3--2) between the Inner and Outer Galaxy by comparing \emph{Herschel}-derived H$_2$ column densities with data from the COHRS \citep{Dempsey+13,Park+23}, CHIMPS \citep{Rigby+16}, and PAMS data, summarising the results in Tables~\ref{tab:X12CO} and \ref{tab:X13CO}. The Inner Galaxy sight-line used was at $\ell=30$\degr, whose emission is dominated by the W43 star-forming region at a distance of 5.2\,kpc \citep{Urquhart+18}, corresponding to a Galactocentric radius of $\rgc = 4.5$\,kpc. We assume that our Inner Galaxy $X$-factors are weighted somewhere slightly closer along the line of sight than W43, so we will take $\rgc = 5.0$\,kpc as the representative value for the Inner Galaxy results\footnote{The intensity-weighted Galactocentric radius from the \citet{Rigby+19} catalogue of clumps in the longitude range of the tile gives a compatible result.}, while the PAMS regions are located close to $\rgc = 9.5$\,kpc.

We found that the distributions of \xtwco\ and \xtco\ values calculated on a pixel-by-pixel basis were significantly different between the Inner and Outer Galaxy regions (and this is confirmed by both Kolmogorov-Smirnov and Anderson-Darling statistical tests). For \twco, the difference is relatively small. Our suggested global value of \xtwco = \xtwcounresolved\ is weighted towards the Inner Galaxy value, and is identical to the value adopted by \citet{Colombo+19} who also used COHRS data. If one considers a common formulation of
\begin{equation} \label{eq:R31}
X_{^{12}\mathrm{CO (3-2})} = \frac{X_{^{12}\mathrm{CO(1-0)}}}{R_{31}},
\end{equation}
\noindent where $R_{31}$ is the \twco\ (3--2)/(1--0) line ratio, these results imply an average value of $R_{31}=0.5$, which is consistent with observations of nearby galaxies \citep[e.g][]{Lamperti+20}, high-redshift galaxies \citep[e.g.][]{Aravena+14} and simulations \citep[e.g.][]{Penaloza+18}.
The Inner and Outer Galaxy distributions of \xtwco\ values in Fig.~\ref{fig:XCOhist} overlap substantially, and representative values for the Outer Galaxy tend to be a factor of 1.3--2.0 lower than their Inner Galaxy counterparts (see Table~\ref{tab:X12CO}). Taking equation (\ref{eq:R31}) literally, these results would imply that $R_{31}$ increases with Galactocentric distance, from $\sim$0.5 to 0.7 between 5.0 and 9.5\,kpc based on equation (ref{eq:R31}). This is contrary to the shallow gradients in $R_{21}$ (which we assume would hold the same trend) found by \citet{Sakamoto+97} in the Milky Way and \citet{DenBrok+22} in M51, but this implication would only hold if the values of \xtwco\ and \xco\ do not vary at the same time. Since we are not directly probing the intensity ratio here, other factors may affect this result.

In equation (\ref{eq:coldens}) in Section~\ref{sec:XCO} we used a single value of $\gamma = 100$ for the gas-to-dust mass ratio, which was shown by \citet{Giannetti+17} to have a Galactocentric dependence (with a large scatter). The Galactic gradient presented in that study would imply a variation of $\gamma=75$ to 185 between $\rgc=$5.0 to 9.5\,kpc, and such a variation could reverse the apparent increasing trend in $R_{31}$ to a decreasing one, and simultaneously reduce the magnitude of the trend. Rescaling the Inner and Outer Galaxy \xco\ values based on this $\gamma$ gradient would result in \xtwco $\approx 3.0 \times 10^{20}$ and $4.5\times10^{20}$\,\xunits\ for the Inner and Outer Galaxy, respectively, implying $R_{31}$ decreasing from $\sim$0.65 to 0.45 over the same range. For \tco\ (3--2), the trend also reverses, resulting in \xtco $\approx 3.6 \times 10^{21}$ and $4.4 \times 10^{21}$\,\xunits\ for the Inner and Outer Galaxy, respectively. Considering this variation, a value of $4.0\times 10^{21}$\,\xunits\ is appropriate for general use in the Galactic disc for sources lying within $\sim$4--10\,kpc of the Galactic Centre, and incorporates the $\gamma$ gradient. A factor of 1.5 uncertainty encapsulates much of the variation we have seen.

Our Inner and Outer Galaxy values of \xtwco\ and \xtco\ do not vary significantly, given the uncertainties in the gas-to-dust mass ratio, and our results are not inconsistent with a radially decreasing $R_{31}$. More-robust direct measurements of $R_{31}$ could be obtained in future at $\sim$50-arcseconds resolution using the MWISP data, or at $\sim$20-arcsec resolution using a combination of CHIMPS, PAMS, COHRS, and FUGIN data.

In all cases, the relationship between H$_2$ column density and integrated intensity of \twco\ and \tco\ (3--2) emission is, of course, more complex than a simple multiplicative factor as can clearly be seen in Fig. \ref{fig:XCO} Panels a) and d). \citet{Barnes+15} remarked on similar behaviour for \twco\ (1--0) emission and suggested adopting a power-law relationship for the column density of CO where $N(\mathrm{CO}) \propto W(^{12}\mathrm{CO}\,1\mathrm{-}0)^{1.38}$, however, the relationships that we report are clearly also not linear in log-space,  indicating that even a single power law would not accurately reproduce the behaviour. A power law with an index of $> 1$ would help reproduce the super-linear parts of the distributions in Figs.~\ref{fig:XCO} a) and d), but these would then under-predict $N$(H$_2$) at low CO intensity. Our Method ii) values that incorporate a background column density could help alleviate this issue, but we cannot be certain whether the background column density is a result of the greater column of Galactic disc that the Inner Galaxy sight-line contains, or if it is reflective of a difference in excitation conditions of CO; sub-thermal emission of CO would also cause an flattening of the relationship at low CO intensity. We therefore recommend that the single representative \xco\ values should be used in the knowledge that more-accurate column densities may be determined through LTE modelling of the combination of \twco, \tco, and -- where available -- \ceo\ emission in combination. 

\subsection{Galactocentric gradients in molecular-cloud properties}

In Section~\ref{sec:properties}, we compared the properties of sources extracted from PAMS with those from a different Inner Galaxy sample from CHIMPS. Here, we created a distance-limited sample of CHIMPS clouds with distances between 2--4\,kpc -- approximately matching the PAMS range of distances -- corresponding to Galactocentric radii of 5--7\,kpc. In Fig.~\ref{fig:radius} we found no substantial differences between two types of radius measurement between the two samples, when considering the \fwchimps\ sources extraction that was designed to be replicate the CHIMPS setup as closely as possible, indicating that the emission profiles of sources in the two samples are similar. The size--linewidth and mass--radius relationships of the two samples of extracted sources sources in Fig.~\ref{fig:scaling} were again found to be consistent. 

One notable difference is that the masses of PAMS sources tend to be greater for a given size-scale (Fig.~\ref{fig:scaling}), with difference between the distributions of $\log_{10}(M/\mathrm{M}_\odot)$ of $\sim$0.15 dex (i.e. around 40 per cent). A surprising result follows, which is that a greater proportion of the PAMS sample satisfy the \citet{Kauffmann+Pillai10} criterion for HMSF than for CHIMPS, with 96 and 63 per cent of the fitted samples doing so, respectively. At face value, this would suggest that the molecular clouds at $\rgc = 9.5$\,kpc have more massive star-forming potential than those at $\rgc \sim 6.0$\,kpc, which is contrary to our expectations, e.g. the decreasing fraction of star-forming Hi-GAL clumps reported by \citet{Ragan+16}. Since the \fwchimps\ source extraction accounts for the differences in data quality between PAMS and CHIMPS, we can rule out biases arising from differences in sensitivity. This discrepancy is made worse if we account for the Galactic gradient in the gas-to-dust mass ratio $\gamma$, discussed in the previous section, which would suggest that our Inner and Outer Galaxy \xtco\ factors were over- and under-estimated, respectively. On the other hand, \citet{Giannetti+17} point to large uncertainties on $\gamma$ for individual star-forming regions, as well as intrinsic scatter, and so this may partly be a result of the small number of independent star-forming complexes that PAMS covers. Additionally, while the Outer Galaxy \xtco\ values derived in Section~\ref{sec:XCO} for the various methods varied little, the Inner Galaxy values were wider-ranging, and the Method iii) value is $\sim$50 per cent larger than the Method i) value we adopted, and this may account for another part of the discrepancy. Furthermore, selection biases resulting from the PAMS survey being targeted towards well-known star-forming regions (in contrast to the blind survey mode of CHIMPS) may account for a further part of the discrepancy. While we can find no single explanation that is entirely satisfactory, we suggest a combination of the above factors may explain this result. Overall, we do not find any significant differences between the sources in the Inner and Outer Galaxy over $5 \leq \rgc \leq 10$\,kpc.

\section{Conclusions} \label{sec:conclusions}

 We have presented the Perseus Arm Molecular Survey (PAMS), a survey of \twco, \tco, and \ceo\ (3--2), covering $\sim$8 deg$^2$ over several molecular-cloud complexes in the Outer Galaxy, at Galactocentric radii of $\sim$9.5 kpc. In Section~\ref{sec:XCO}, we calculated \xco\ factors for \twco\ and \tco\ (3--2), which convert the integrated intensity of the CO emission into molecular-hydrogen column density, and examined the distributions of pixel values. In Section \ref{sec:properties}, we compared basic properties of sources extracted from PAMS with equivalent values from the CHIMPS survey in order to probe any differences between these Inner and Outer Galaxy environments, covering a range of Galactocentric radii between 4--10\,kpc.

Our main findings are as follows:
\begin{enumerate}
    \item The systematic variation in \xco\ values derived the for Inner and Outer Galaxy were generally small compared with the variation arising from the different methods, as well as the spread in individual pixel values, but that difference was stronger in \tco\ (3--2) than \twco\ (3--2). 
    \item We recommend the usage of a value of \xtwco=\xtwcounresolved\ to convert \twco\ (3--2) integrated intensity to molecular-hydrogen column density, with a factor of 1.5 uncertainty, in agreement with previous studies \citep[e.g.][]{Colombo+19}.
    \item For \tco\ (3--2), we recommend the usage of a value of \xtco$\,= 4.0 \times 10^{21}$\,\xunits\ to convert \tco\ (3--2) integrated intensity to molecular-hydrogen column density, with a factor of 1.5 uncertainty.
    \item Although the \xtco\ values we recovered are significantly different for the Inner and Outer Galaxy (traced by CHIMPS and PAMS), with \xtco$\,=4.8 \times 10^{21}$\,\xunits\ for the Inner Galaxy and $2.4 \times 10^{21}$\,\xunits\ for the Outer Galaxy, accounting for the Galactic gradient in gas-to-dust mass ratio can resolve much of the difference. Calculating H$_2$ column densities from LTE analysis of co-spatial \twco\ and \tco\ emission would be preferable to relying on \xco\ factors. 
    \item We did not find any significant differences between the emission profiles, of sources extracted from the \tco\ (3--2) PAMS data at Galactocentric distances of 9--10\,kpc and from an equivalent sample of sources extracted from CHIMPS \tco\ (3--2) data at Galactocentric distances of 5--7\,kpc. Similarly, the size--linewidth and mass--radius relationships were also compatible.
    \item Although the distributions of masses of PAMS Outer Galaxy sources are shifted to greater values than their CHIMPS Inner Galaxy counterparts for a given size scale, most of the $\sim$0.2 dex difference can be explained by a combination of variations in gas-to-dust mass ratios and selection biases.
\end{enumerate}

Finally, we have demonstrated that the PAMS data are a valuable addition to the existing repertoire of publicly available \twco, \tco, and \ceo\ (3--2) survey data in the Outer Galaxy. In combination with surveys such as CHIMPS, CHIMPS2, and COHRS, these data extend the baseline in Galactocentric radius in what can be studied in the 3--2 rotational transition across different Galactic environments, in addition to the growing number of surveys covering 2--1 and 1--0.

\section*{Acknowledgements}

We would like to thank the anonymous referee whose review helped to improve the quality and accuracy of this manuscript. We thank Andrea Giannetti for a thoughtful discussion about gas-to-dust mass ratios. AJR acknowledges postdoctoral support from the University of Leeds. The JCMT is operated by the East Asian Observatory on behalf of The National Astronomical Observatory of Japan; Academia Sinica Institute of Astronomy and Astrophysics; the Korea Astronomy and Space Science Institute; the National Astronomical Research Institute of Thailand; Center for Astronomical Mega-Science (as well as the National Key R\&D Program of China with No. 2017YFA0402700). Additional funding support is provided by the Science and Technology Facilities Council of the United Kingdom and participating universities and organizations in the United Kingdom and Canada. The JCMT has historically been operated by the Joint Astronomy Centre on behalf of the Science and Technology Facilities Council of the United Kingdom, the National Research Council of Canada and the Netherlands Organisation for Scientific Research. The authors wish to recognize and acknowledge the very significant cultural role and reverence that the summit of Maunakea has always had within the indigenous Hawaiian community.  We are most fortunate to have the opportunity to conduct observations from this mountain. This research used the facilities of the Canadian Astronomy Data Centre operated by the National Research Council of Canada with the support of the Canadian Space Agency. This research has made use of NASA's Astrophysics Data System Bibliographic Services'.

\noindent \textbf{Software} \\
\href{https://www.python.org/}{Python} packages:
\href{https://www.astropy.org/}{\texttt{astropy}} \citep{TheAstropyCollaboration22},
\href{https://ipython.org/}{\texttt{ipython}} \citep{Perez+Granger07},
\href{https://matplotlib.org/}{\texttt{matplotlib}} \citep{Hunter07},
\href{https://github.com/pjcigan/multicolorfits}{\texttt{multicolorfits}} \citep{Cigan19},
\href{https://numpy.org/}{\texttt{numpy}} \citep{Harris+20},
\href{https://scipy.org/}{\texttt{scipy}} \citep{Virtanen+20},
\href{https://scikit-learn.org/stable/}{\texttt{scikit-learn}} \citep{Pedregosa+11}.

\href{https://starlink.eao.hawaii.edu/starlink}{Starlink} \citep{Currie+14} applications:
\href{https://starlink.eao.hawaii.edu/docs/sun214.htx/sun214.html}{\texttt{gaia}},
\href{https://starlink.eao.hawaii.edu/docs/sun95.htx/sun95.html}{\texttt{kappa}} \citep{Currie+Berry13},
\href{http://starlink.eao.hawaii.edu/docs/sun255.htx/sun255.html#toc}{\texttt{cupid/fellwalker}} \citep{Berry15},
\href{http://starlink.eao.hawaii.edu/docs/sun260.htx/sun260.html}{\texttt{orac-dr}} \citep{Jenness+15}.

Other software:
\href{https://www.star.bris.ac.uk/~mbt/topcat/}{\texttt{topcat}} \citep{Taylor05}.

%%%%%%%%%%%%%%%%%%%%%%%%%%%%%%%%%%%%%%%%%%%%%%%%%%
\section*{Data Availability}

We make the PAMS data publicly available at \red{\url{https://dx.doi.org/10.11570/25.0001}}. This repository includes the \twco, \tco, and \ceo\ (3--2) mosaics of each region, and the \fw\ catalogues with their corresponding assignment masks.

%%%%%%%%%%%%%%%%%%%% REFERENCES %%%%%%%%%%%%%%%%%%

\bibliographystyle{mnras}
\bibliography{PAMS} % if your bibtex file is called example.bib

%%%%%%%%%%%%%%%%%%%%%%%%%%%%%%%%%%%%%%%%%%%%%%%%%%

%%%%%%%%%%%%%%%%% APPENDICES %%%%%%%%%%%%%%%%%%%%%

\appendix

\section{ORAC-DR Recipe Parameters} \label{app:recpars}

In this Appendix, we list of several sections of the recipe parameters used for the data-reduction recipe \textsc{reduce\_science\_narrowline} in {\sc orac-dr}. The first block relates to the pixel size and the binning method:

\begin{verbatim}
[REDUCE_SCIENCE_NARROWLINE]                                                     
#                                                                               
# MAKECUBE parameters                                                           
PIXEL_SCALE = 6.0                                                               
SPREAD_METHOD = Gauss                                                           
SPREAD_WIDTH = 8                                                                
SPREAD_FWHM_OR_ZERO = 6                                                         
#                                                                               
REBIN = 0.3                                                                     
#                            
\end{verbatim}
which specifies the use of 6.0-arcsecond-wide pixels, and the use of a Gaussian smoothing kernel with FWHM of 8 arcsec to assist with the binning of pixel values onto the new pixel grid. The spreading function is curtailed at 6 arcsec, as specified by \verb|SPREAD_FWHM_OR_ZERO|. Finally, the \verb|REBIN| parameter specifies that the cube will be regridded to a 0.3\,\kms-wide velocity channels.

\begin{verbatim}
# Tiling and chunking                                                           
TILE = 0                                                                        
CHUNKSIZE = 12288                                                               
CUBE_MAXSIZE = 1536   
#
\end{verbatim}
This block specifies that the entire cube should be treated as a single observation, and not be broken up into smaller tiles for memory-saving reasons. This produces a more convenient output.

\begin{verbatim}                                                   
# Baseline                                                                      
BASELINE_ORDER = 1                                                              
BASELINE_LINEARITY = 1                                                          
BASELINE_LINEARITY_LINEWIDTH = -80:-20                                          
\end{verbatim}
The above parameters specify that a first-order polynomial baseline should be used for the fitting, and that the region of $-$80 to $-20$\,\kms\ in the spectrum should be excluded when performing baseline linearity tests for each receptor.

\begin{verbatim}
# # Reference-spectrum removal from timeseries cubes                            
# --- Manual location                                                                                               
SUBTRACT_REF_SPECTRUM = 1                                                      
REF_SPECTRUM_COMBINE_REFPOS = 1                                                
REF_SPECTRUM_REGIONS = -15.0:-11.5,14.0:17.0                         
\end{verbatim}
This final set of parameters was optionally used in instances where off-position absorption was suspected to be present. This often reveals itself as a velocity range showing absorption features in the cube-average spectrum. By enabling \verb|SUBTRACT_REF_SPECTRUM| and \verb|REF_SPECTRUM_COMBINE_REFPOS|, {\sc orac-dr} interpolates the reference spectrum across the velocity range or ranges identified in the \verb|REF_SPECTRUM_REGIONS| setting, which consists of a comma-separated list of regions with suspected off-position emission. In this particular instance, the ranges of $-$15.0 to $-11.5$ and 14.0 to 17.0\,\kms\ were interpolated over in the off- (reference-) position spectrum.

\section{Preview images of the PAMS regions} \label{app:regions}

In this appendix we display integrated intensity (moment 0) images of the PAMS regions in \twco, \tco, and \ceo (3--2), along with the corresponding position-velocity diagrams. We do not include NGC\,7538 which has already been shown in Fig.~\ref{fig:NGC7538}. In all cases, the cubes have been masked using \fw, adopting the \fwlores\ masks discussed in Section~\ref{sec:sourceextraction}. 

\begin{figure*}
    \centering
    \includegraphics[width=\textwidth]{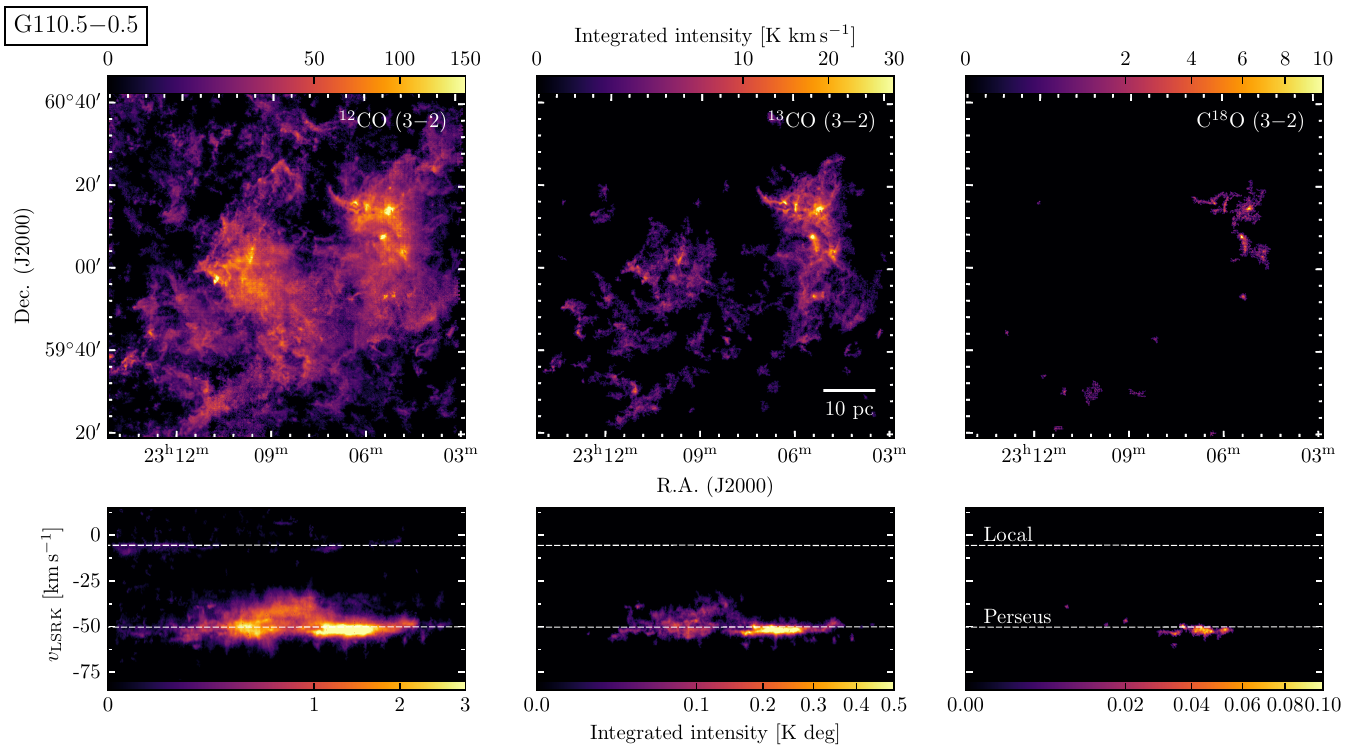}
    \caption{\emph{Top row}: Integrated intensity (moment 0) maps of \twco, \tco, and \ceo\ (3--2) emission of G110.5$+$0.5. \emph{Bottom row}: Position-velocity maps of the above maps integrated over declination.}
    \label{fig:appendix_G110}
\end{figure*}

\begin{figure*}
    \centering
    \includegraphics[width=\textwidth]{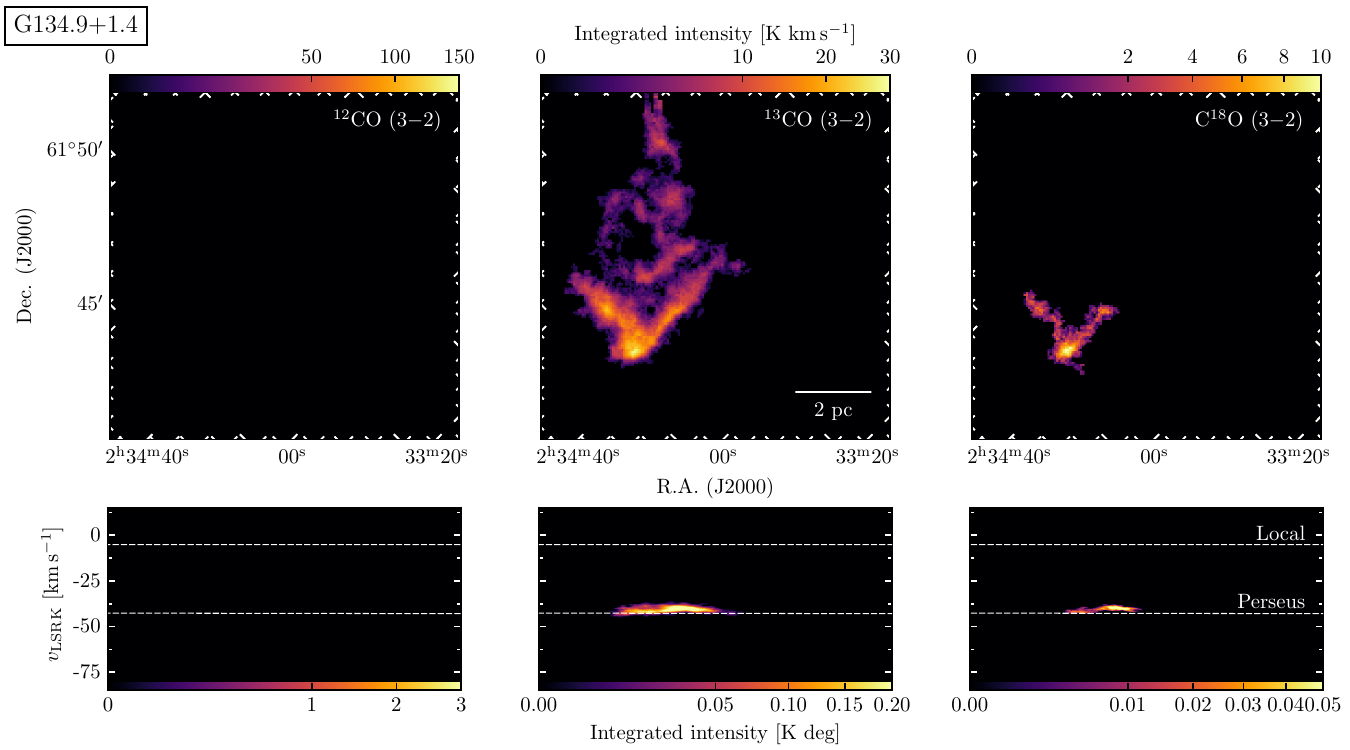}
    \caption{\emph{Top row}: Integrated intensity (moment 0) maps of \twco, \tco, and \ceo\ (3--2) emission of G134.9$+$1.4. \emph{Bottom row}: Position-velocity maps of the above maps integrated over the y-axis. There are no \twco\ (3--2) data for this region.}
    \label{fig:appendix_G134}
\end{figure*}

\begin{figure*}
    \centering
    \includegraphics[width=\textwidth]{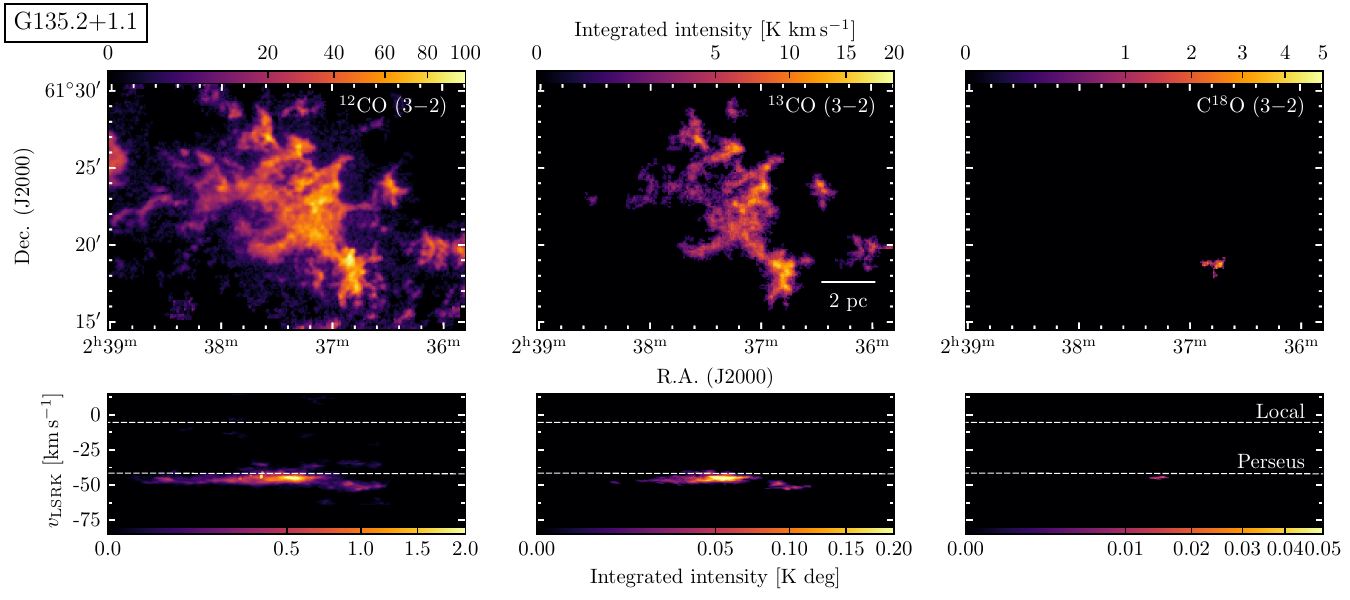}
    \caption{\emph{Top row}: Integrated intensity (moment 0) maps of \tco, and \ceo\ (3--2) emission of G135.2$+$1.1. \emph{Bottom row}: Position-velocity maps of the above maps integrated over declination.}
    \label{fig:appendix_G135}
\end{figure*}

\begin{figure*}
    \centering
    \includegraphics[width=\textwidth]{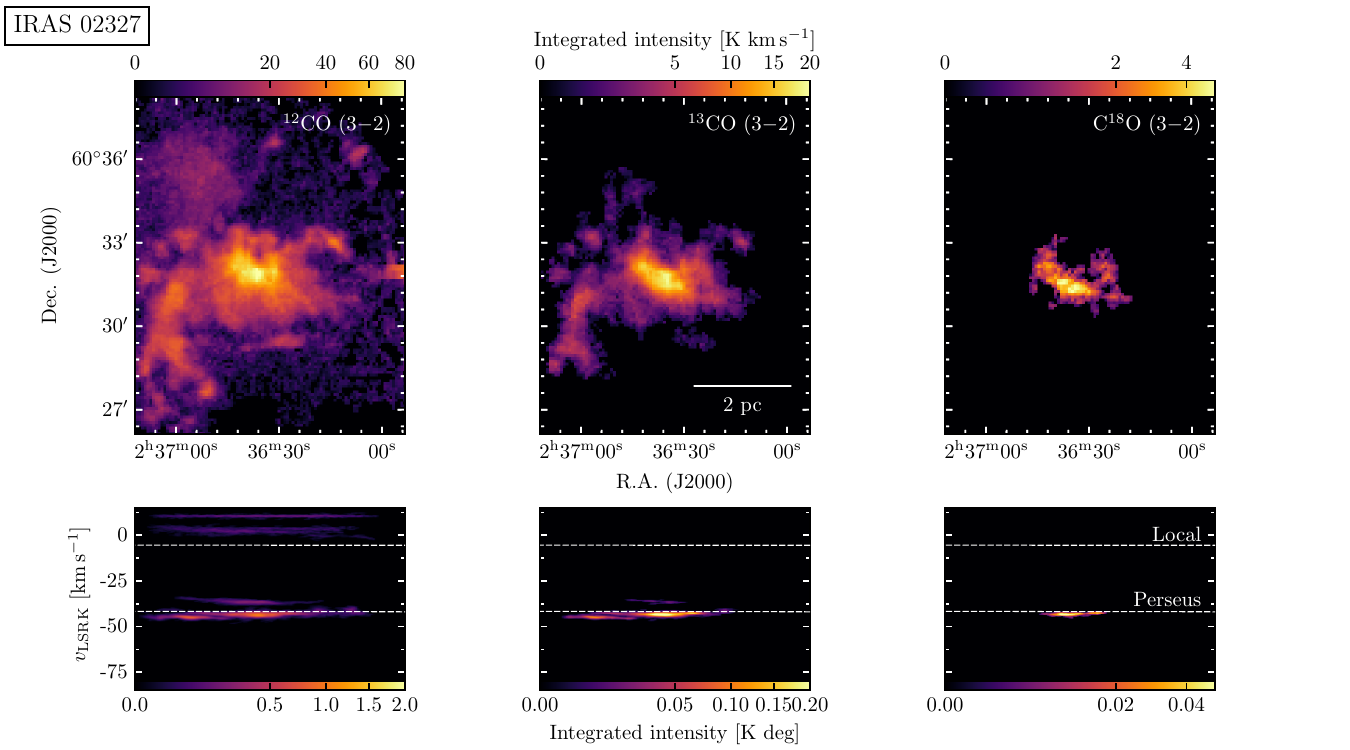}
    \caption{\emph{Top row}: Integrated intensity (moment 0) maps of \twco, \tco, and \ceo\ (3--2) emission of G135.2$+$1.1. \emph{Bottom row}: Position-velocity maps of the above maps integrated over declination.}
    \label{fig:appendix_IRAS02327}
\end{figure*}

\begin{figure*}
    \centering
    \includegraphics[width=\textwidth]{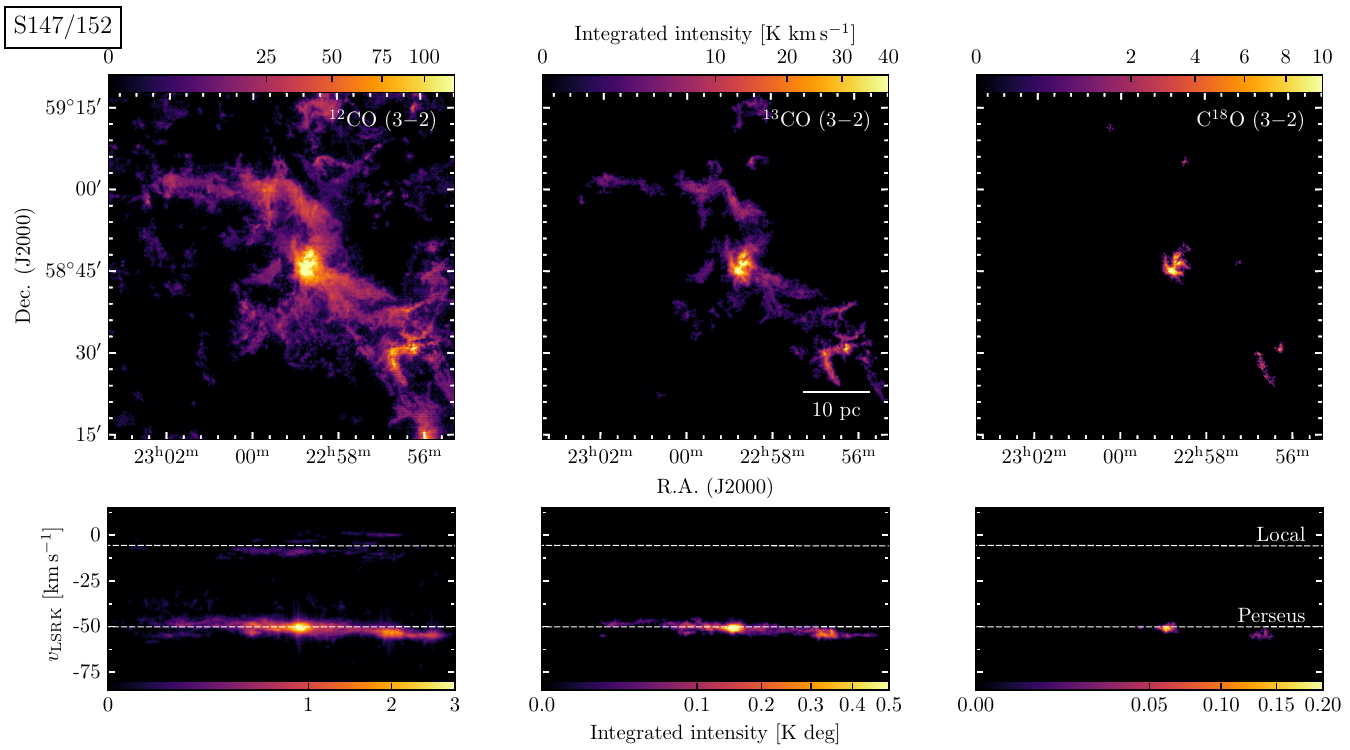}
    \caption{\emph{Top row}: Integrated intensity (moment 0) maps of \twco, \tco, and \ceo\ (3--2) emission of S147/152. \emph{Bottom row}: Position-velocity maps of the above maps integrated over declination.}
    \label{fig:appendix_S152}
\end{figure*}

\begin{figure*}
    \centering
    \includegraphics[width=\textwidth]{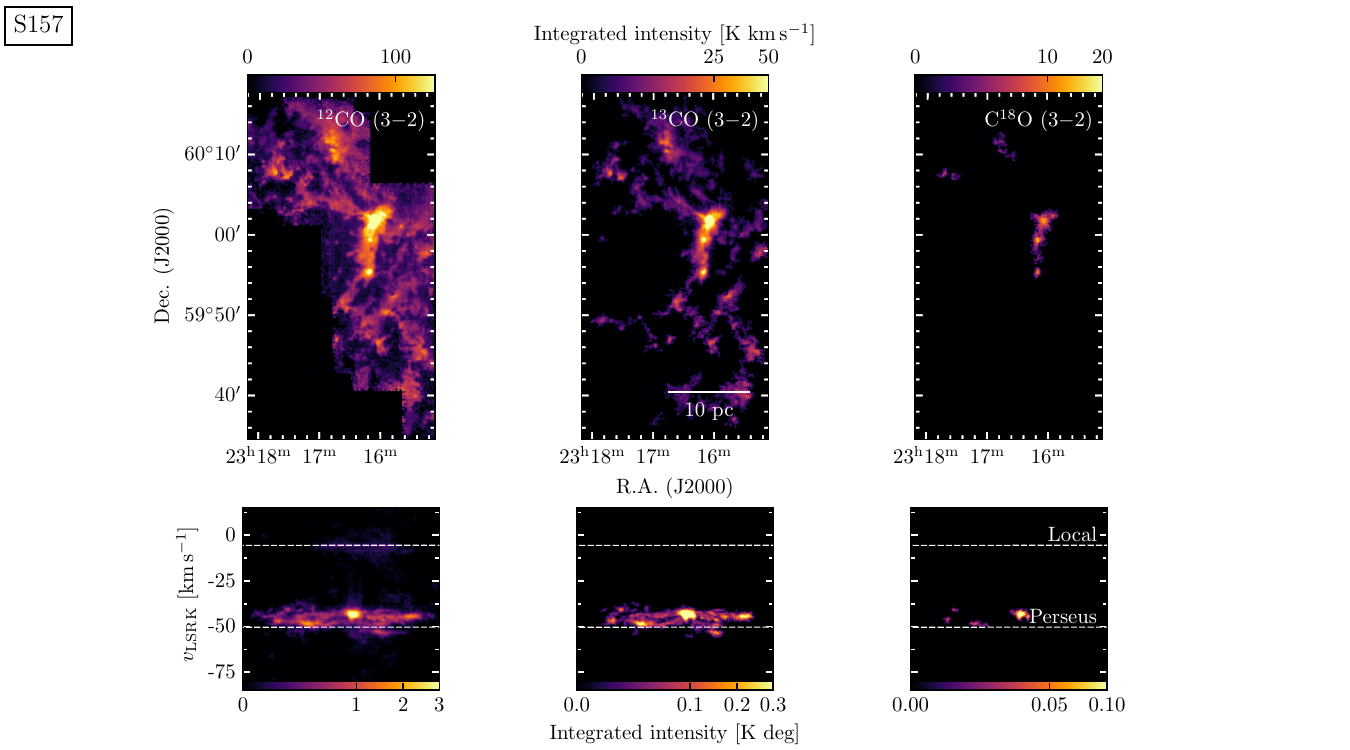}
    \caption{\emph{Top row}: Integrated intensity (moment 0) maps of \twco, \tco, and \ceo\ (3--2) emission of S157. \emph{Bottom row}: Position-velocity maps of the above maps integrated over declination.}
    \label{fig:appendix_S157}
\end{figure*}

\begin{figure*}
    \centering
    \includegraphics[width=\textwidth]{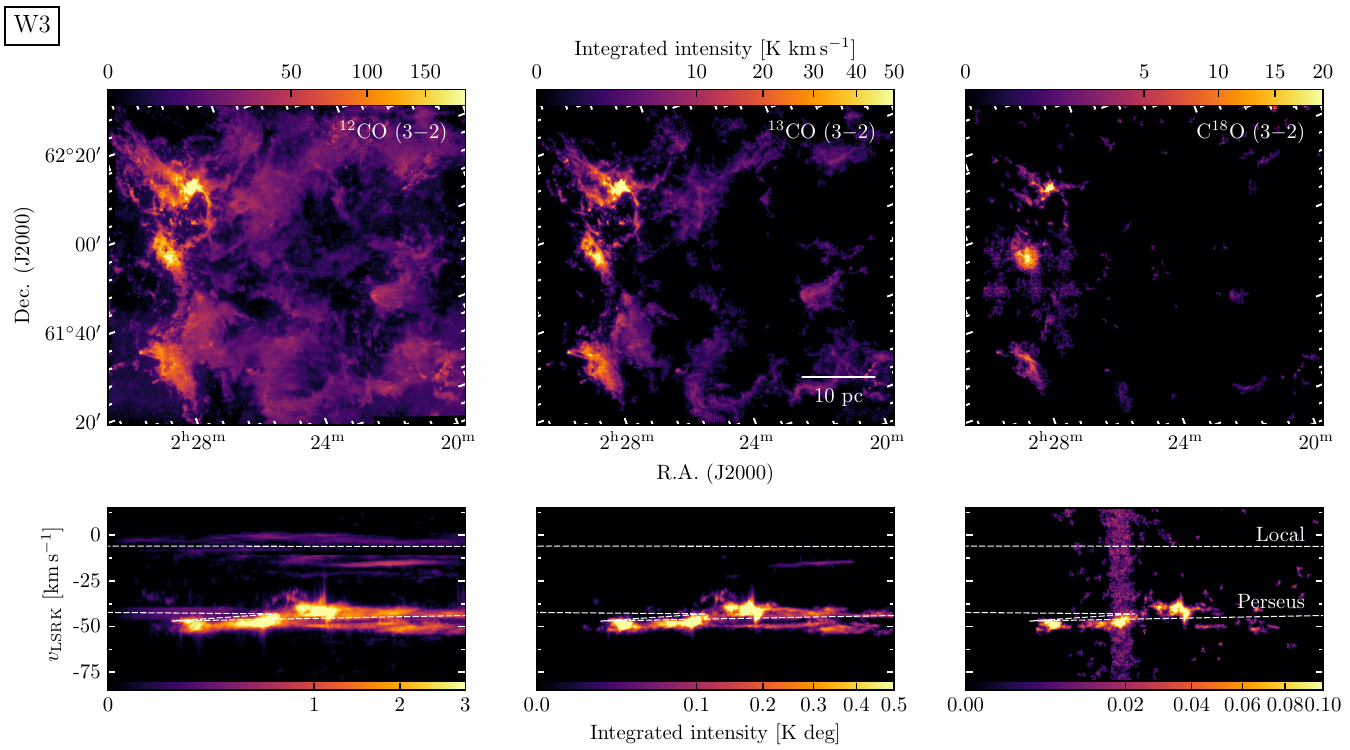}
    \caption{\emph{Top row}: Integrated intensity (moment 0) maps of \twco, \tco, and \ceo\ (3--2) emission of W3. \emph{Bottom row}: Position-velocity maps of the above maps integrated over the y-axis.}
    \label{fig:appendix_W3}
\end{figure*}

\begin{figure*}
    \centering
    \includegraphics[width=\textwidth]{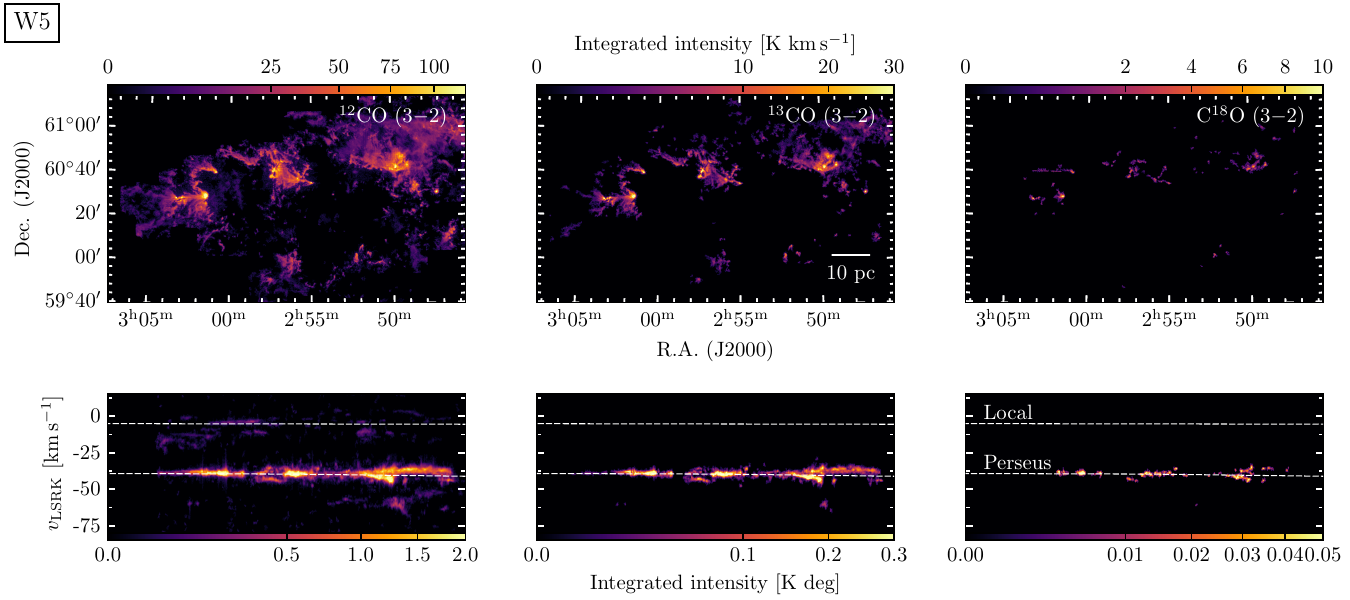}
    \caption{\emph{Top row}: Integrated intensity (moment 0) maps of \twco, \tco, and \ceo\ (3--2) emission of W5. \emph{Bottom row}: Position-velocity maps of the above maps integrated over the y-axis.}
    \label{fig:appendix_W5}
\end{figure*}

\clearpage
\balance
\section{\fw\ configuration} \label{app:fwconfig}

In Section~\ref{sec:sourceextraction} we described our usage of \fw\ to extract sources from the \tco\ (3--2) PAMS data. Our parameter setup for the \fwhires\ extraction was as follows.
\begin{verbatim}                                                   
    FellWalker.AllowEdge=1
    FellWalker.CleanIter=0
    FellWalker.FlatSlope=1*RMS
    FellWalker.FwhmBeam=3
    FellWalker.MaxBad=0.05
    FellWalker.MinDip=5*RMS
    FellWalker.MinHeight=3*RMS
    FellWalker.MinPix=16
    FellWalker.MaxJump=0
    FellWalker.Noise=1*RMS
    FellWalker.RMS=1
    FellWalker.VeloRes=1
\end{verbatim}
The parameter selection for the \fwlores\ extraction was identical, with the following exception.
\begin{verbatim}
    FellWalker.MinDip=1000*RMS
\end{verbatim}
\noindent This change effectively suppresses the ability of \fw\ to identify substructures within isolated islands of emission, and thus will recover the largest-possible complexes of contiguous pixels of emission.

The parameters used for the \fwchimps\ extraction were the same as for \fwhires\ with the following alterations.
\begin{verbatim}
    FellWalker.AllowEdge=0
    FellWalker.CleanIter=1
    FellWalker.MaxJump=4
    FellWalker.RMS=1.7
    FellWalker.VeloRes=0
\end{verbatim}
\noindent This setup was chosen to match the extraction of \citet{Rigby+16} as closely as possible. The key difference is that \verb|RMS| is set to 1.7 to reflect higher noise levels in the CHIMPS compared with PAMS data after smoothing to 27.4-arcsec resolution. The parameters \verb|FlatSlope|, \verb|MinDip|, \verb|MinHeight|, and \verb|Noise| are scaled up by the same amount. The differences between the \fwhires\ setup compared with \fwchimps\ were generally selected to combat spurious sources, and especially the archipelago source types discussed briefly in Section~\ref{sec:properties}; setting both \verb|CleanIter| and \verb|MaxJump| to zero, and \verb|VeloRes| to 1 helps suppress these source types. It was also necessary to adopt value of 1 for \verb|AllowEdge| due to the extra sensitivity and targeted mapping of the PAMS observations, which meant that emission was often detected up to the edges of the image, and a zero value would have excluded significant sources.

% \balance
\section{\fw\ catalogue descriptions} \label{app:table}

The columns featuring in the \fwhires, \fwlores, and \fwchimps\ catalogues are listed Table~\ref{tab:columndescriptions}. The only difference between the three is that the \fwhires\ catalogue features an additional \verb|Parent_ID| column, which allows the \fwhires\ sources to be connected to their parent larger-scale structure in the \fwlores\ extraction.

\begin{table*}
    \centering
    \caption{Column descriptions for the provided catalogues,}
    \begin{tabular}{cllll}
    \hline
    Column number & Name & Data type & Units & Description \\
    \hline
    1 & \verb|Region| & string & None &  PAMS region \\
    2 & \verb|ID| & integer & None &  Voxel ID in mask \\
    3 & \verb|RA_cen| & float & deg &  Centroid RA coordinate \\
    4 & \verb|Dec_cen| & float & deg &  Centroid Dec coordinate  \\
    5 & \verb|Vlsr_cen| & float & \kms &  Centroid \vlsr\ coordinate \\
    6 & \verb|RA_peak| & float & deg &  Peak RA coordinate \\
    7 & \verb|Dec_peak| & float & deg &  Peak Dec coordinate \\
    8 & \verb|Vlsr_peak| & float & \kms &  Peak \vlsr\ coordinate \\
    9 & \verb|Sigma_RA| & float & arcsec &  Intensity-weighted dispersion in RA \\
    10 & \verb|Sigma_Dec| & float & arcsec &  Intensity-weighted dispersion in Dec \\
    11 & \verb|Sigma_Vlsr| & float & \kms &  Intensity-weighted dispersion in \vlsr\ \\
    12 & \verb|Sigma_Vlsr_corrected| & float & \kms &  Estimated linewidth corrected for SNR \\
    13 & \verb|Sigma_Vlsr_err| & float & \kms &  Uncertainty on linewidth \\
    14 & \verb|Sum| & float & K &  Sum of \tco\ (3--2) voxel values \\
    15 & \verb|Peak| & float & K &  Peak \tco\ (3--2) voxel value \\
    16 & \verb|major_sigma| & float & deg &  Semi-major axis of ellipse \\
    17 & \verb|minor_sigma| & float & deg &  Semi-minor axis of ellipse \\
    18 & \verb|position_angle| & float & deg &  Position angle of ellipse (E of N) \\
    19 & \verb|n_pixels| & integer & None &  Number of pixels in projected area \\
    20 & \verb|n_voxels| & integer & None &  Number of voxels in source \\
    21 & \verb|polygon| & string & None &  (RA, Dec) vertices of polygon describing source \\
    22 & \verb|RA_cen_snr| & float & deg &  Centroid RA coordinate in SNR cube \\
    23 & \verb|Dec_cen_snr| & float & deg &  Centroid Dec coordinate in SNR cube \\
    24 & \verb|Vlsr_cen_snr| & float & \kms &  Centroid \vlsr\ coordinate in SNR cube \\
    25 & \verb|RA_peak_snr| & float & deg &  Peak RA coordinate in SNR cube \\
    26 & \verb|Dec_peak_snr| & float & deg &  Peak Dec coordinate in SNR cube \\
    27 & \verb|Vlsr_peak_snr| & float & \kms &  Peak \vlsr\ coordinate in SNR cube \\
    28 & \verb|Sigma_RA_snr| & float & arcsec &  Intensity-weighted dispersion in RA in SNR cube \\
    29 & \verb|Sigma_Dec_snr| & float & arcsec &  Intensity-weighted dispersion in Dec in SNR cube \\
    30 & \verb|Sigma_Vlsr_snr| & float & \kms &  Intensity-weighted dispersion of \vlsr\ in SNR cube \\
    31 & \verb|Sum_snr| & float & None &  Sum of voxel values in SNR cube \\
    32 & \verb|Peak_snr| & float & None &  Peak voxel value in SNR cube \\
    33 & \verb|dkpc| & float & kpc &  Heliocentric distance to region \\
    34 & \verb|derr| & float & kpc &  Error on heliocentric distance to region \\
    35 & \verb|Rgc| & float & kpc &  Galactocentric radius of region \\
    36 & \verb|R_eq_arcsec_nodec| & float & arcsec &  Radius of circle with equivalent area \\
    37 & \verb|R_eq_arcsec| & float & arcsec &  Beam-deconvolved radius of circle with equivalent area \\
    38 & \verb|R_eq_pc| & float & pc &  Beam-deconvolved radius of circle with equivalent area \\
    39 & \verb|R_eq_pc_err| & float & pc &  Error on \verb|R_eq_pc| \\
    40 & \verb|R_sig_arcsec_nodec| & float & arcsec &  Intensity-weighted radius \\
    41 & \verb|R_sig_arcsec| & float & arcsec &  Beam-deconvolved intensity-weighted radius \\
    42 & \verb|R_sig_pc| & float & pc &  Beam-deconvolved intensity-weighted radius \\
    43 & \verb|R_sig_pc_err| & float & pc &  Error on \verb|R_sig_pc| \\
    44 & \verb|Mass_Xco| & float & \msun &  Mass estimated using \tco\ (3--2) X-factor \\
    45 & \verb|Mass_Xco_lo| & float & \msun &  16th percentile of \verb|Mass_Xco| distribution \\
    46 & \verb|Mass_Xco_up| & float & \msun &  84th percentile of \verb|Mass_Xco| distribution \\
    47* & \verb|Parent_ID| & integer & None &  ID of parent source in \verb|fwlores| catalogue\\
    \hline
    & & & & * (\verb|fwhires| catalogue only) \\
    \end{tabular}
    \label{tab:columndescriptions}
\end{table*}

%%%%%%%%%%%%%%%%%%%%%%%%%%%%%%%%%%%%%%%%%%%%%%%%%%

% Don't change these lines
\bsp	% typesetting comment
\label{lastpage}
\end{document}